\documentclass[aps,11pt,superscriptaddress,longbibliography]{revtex4-1}
\usepackage{amsmath, amsfonts,amssymb}
\def\*#1{\mathbf{#1}}

\usepackage{float}
\usepackage{graphicx}

\usepackage{hyperref}
\usepackage{comment}
\usepackage{color}
\usepackage{subfigure}

\usepackage{textpos}

\begin{document}

\begin{textblock*}{10cm}(0.5cm,-1.2cm)
  \footnotesize Published in the journal Physical Review Fluids.
\end{textblock*}

\title{Immiscible Rayleigh-Taylor turbulence \\ using mesoscopic lattice Boltzmann algorithms}
\author{H.S. Tavares}\email{hugoczpb@impa.br}
\affiliation{Instituto de Matem\'atica Pura e Aplicada -- IMPA, Rio de Janeiro, Brazil}
\author{L. Biferale}
\affiliation{Dept. Physics and INFN, University of Rome Tor Vergata, Italy}
\author{M. Sbragaglia}
\affiliation{Dept. Physics and INFN, University of Rome Tor Vergata, Italy}
\author{A.A. Mailybaev}
\affiliation{Instituto de Matem\'atica Pura e Aplicada -- IMPA, Rio de Janeiro, Brazil}

    \begin{abstract}
    \begin{center}
      (Received 3 September 2020; accepted 29 March 2021; published 13 May 2021)
    \end{center}
   
    
    We studied turbulence induced by the  Rayleigh-Taylor (RT) instability for
	2D immiscible two-component flows by using a multicomponent  lattice
	Boltzmann method with a  Shan-Chen pseudopotential  implemented on graphics processing units. We compare our results with the extension to the 2D case of the
	phenomenological theory for immiscible 3D RT turbulence  studied by Chertkov and collaborators 
	[Phys. Rev. E 71, 055301
(2005)]. 
	Furthermore, we compared  the growth of the
	mixing layer, typical velocity, average density profiles and enstrophy with the equivalent case but for miscible two-component fluid. In
	both miscible and immiscible cases, the expected quadratic growth of the mixing layer and the linear growth of the typical velocity are observed with close long-time asymptotic prefactors but different initial transients. In the immiscible case,  the enstrophy shows a tendency to grow like $\propto t^{3/2}$, with the highest values of vorticity concentrated close to the interface. In addition, we  investigate the evolution of the typical drop size and the behavior of the total length of the interface in the emulsion-like state, showing the existence of a  power law behavior compatible with our phenomenological predictions. Our results can also be considered as a first validation step to extend the application of the lattice Boltzmann tool to study the 3D immiscible case. 
	
	\vspace{0.3cm}
	\noindent DOI: {\color{blue}\url{https://doi.org/10.1103/PhysRevFluids.6.054606}}
	
	\end{abstract}

\maketitle

   \section{Introduction}
   When a heavy fluid is accelerated against a lighter fluid the so-called Rayleigh-Taylor (RT) instability can develop \cite{lord1900investigation, taylor1950instability}, which eventually leads to a mixing layer with a turbulent motion called Rayleigh-Taylor turbulence. In this process the two fluids seek to reduce the total potential energy of the system~\cite{celani2009phase}. The turbulent regime is relevant in many different contexts, for example, in the understanding of the Earth's climate, in the nuclear fusion process~\cite{petrasso1994rayleigh,burrows2000supernova} and as a key mechanism for thermonuclear flames in some types of supernovae~\cite{zingale2005three,schmidt2006tea}. In the context of classical fluids, the incompressible Rayleigh-Taylor turbulence has important properties~\cite{boffetta2017incompressible}, one of the most important of which is the quadratic growth of the mixing layer width. In some cases, important connections have been found with classical theories of turbulence for simple fluids~\cite{frisch1995turbulence,chertkov2003phenomenology,abarzhi2005rayleigh}.
   
   Physical experiments of the RT instability have shown some challenges due to the difficulty of sustaining an unstable density stratification necessary to set up the appropriate initial conditions for the instability~\cite{celani2009phase,ramaprabhu2004experimental,wilson1972excitatory,Zhou20171}. Despite this limitation, considerable advances in numerical simulations of the Rayleigh-Taylor instability have been verified in the past few decades, especially in the context of systems with miscible fluids~\cite{chertkov2003phenomenology,biferale2010high,celani2006rayleigh,boffetta2017incompressible,biferale2018rayleigh,zhou2017rayleigh}. Only a few works have been dedicated to the immiscible case~\cite{celani2009phase,young2006surface,liang2019direct,brackbill1992continuum,carles2006rayleigh,livescu2004compressibility}, and most of them are devoted to the early stages of the instability with little information about the state of developed turbulence. One of the reasons for this is the highly complicated pattern formed by the interfaces that appear in the immiscible case, originating high gradients and singularities in the solutions, which is a source of challenging numerical instabilities in many numerical methods for multicomponent fluids.  Some works tried to close the dynamics in terms of  effective equations for the interface; see~\cite{abarzhi2020scale,abarzhi2019supernova} for a recent discussion.
   
   With respect to the theoretical aspects of the immiscible RT turbulence, it is only recently that a consistent phenomenological study of the effects of surface tension has been proposed by Chertkov and collaborators~\cite{chertkov2005effects}. It followed the earlier work in Ref.~\cite{chertkov2003phenomenology}, where a phenomenological theory was developed for two- and three-dimensional miscible RT turbulence in the Boussinesq approximation.  Said work considers a three-dimensional(3D) scenario, in which the direct energy cascade happens in a range of scales limited by the mixing layer width (integral scale) and the viscous (Kolmogorov) scale, both dependent on time. 
   In the two dimensional case, the lack of energy and enstrophy cascades leads to the assumption of Bolgiano--Obukhov theory describing the cascade of temperature fluctuations in the inertial range~\cite{bolgiano1959turbulent,obukhov1959influence}. Reference \cite{chertkov2005effects} described the theory of three-dimensional immiscible RT turbulence, studying the effects of surface tension in an emulsionlike state and predicting the rate of growth for the typical drop size. 
   
   In the present paper we extend the phenomenological theory of Ref.~\cite{chertkov2005effects} for two-dimensional immiscible RT turbulence assuming the Boussinesq approximation, which is valid in the limit of small density variations~\cite{landau2013fluid,kundu2001fluid}. This extension includes predictions for the growth of the total length of the interface and the typical drop size. We also provide predictions for the evolution of the enstrophy in the miscible and immiscible cases, which have not been addressed earlier. 
  These predictions are tested using numerical simulations based on the multicomponent lattice-Boltzmann method with the Shan-Chen pseudopotential model~\cite{kruger2017lattice,succi2018lattice}. In the immiscible case, this method is able to accurately overcome the inherent numerical complexity caused by the complicated structure of the interface that appears in the fully developed turbulent regime \cite{scarbolo2013unified,celani2009phase,young2006surface}. This method also allows parallel implementations in many situations, which is very important for statistical analyses that requires a substantial number of simulations, as in our numerical verification for the phenomenological predictions. We run several parallel simulations of RT turbulence on graphics processing units (GPUs) using CUDA with a computational grid of resolution $10 000 \times 5000$.
   
   This paper is organized as follows. Section II describes the basic equations for the classical Rayleigh-Taylor system, miscible and immiscible, characterizing the Boussinesq approximation and including the surface tension effects. 
   In Sec. III we describe the multicomponent lattice-Boltzmann method with the Shan-Chen pseudopotential model, and show how to approach the Boussinesq approximation with this method.
      In Sec. IV we construct phenomenological predictions for the mixing layer, typical velocity and averaged density profile, with the respective numerical verifications, showing a direct comparison between the miscible and immiscible cases. Section V is dedicated to the phenomenological properties of the interface. In the first part of that section, we investigate the evolution of the typical drop size and total length of the interface in the emulsion-like state, and at the end we study the evolution of  enstrophy. The statistics for the enstrophy are also used to understand the influence of the interface on small-scale statistics and to verify the validity of the assumption of the Bolgiano-Obukhov regime in our phenomenology. Section VI provides some conclusions and perspectives.

\section{Immiscible and miscible Rayleigh-Taylor systems}

An interface between two fluids of different densities becomes unstable when a heavier fluid is placed above a lighter fluid under gravity~\cite{chandrasekhar2013hydrodynamic}. In the classical formulation of fluid dynamics, the flow is described by the incompressible Navier--Stokes equations
	\begin{equation}
	\label{eq2_1}
	\rho\left( \frac{\partial \mathbf{u}}{\partial t}+\mathbf{u} \cdot \nabla\mathbf{u} \right)
	=-\nabla p+\nabla \cdot \left\lbrace  \mu \left(\nabla \mathbf{u}  +(\nabla \mathbf{u})^T\right) \right\rbrace 
	+\mathbf{f}, \quad \nabla \cdot \mathbf{u} = 0,
	\end{equation}
where $\mathbf{u}$ is the fluid velocity depending on spatial coordinates $\mathbf{x}$ and time $t$, $p$ is the pressure, and $\rho$ and $\mu$ are the fluid density and dynamic viscosity. The buoyancy forcing term is $\mathbf{f} = \rho \mathbf{g}$ with the acceleration of gravity $\mathbf{g}$. In this work, we study two-dimensional flows with $\mathbf{x} = (x,y)$ for two different physical models describing the immiscible and miscible flows.

The immiscible formulation considers two fluid phases with constant densities and viscosities: $\rho_1$ and $\mu_1$ for the first phase and $\rho_2$ and $\mu_2$ for the second phase. We assume that $\rho_1 > \rho_2$, i.e., the first phase is heavier. The  two subdomains occupied by each phase are separated by a moving interface $\Gamma(t)$. Equations of motion for each phase are given by (\ref{eq2_1}) with the corresponding constant values of density and viscosity. At the interface, the boundary conditions take the form
	\begin{equation}
	\label{eq2_2}
	\mathbf{x} \in \Gamma:\quad  
	[\mathbf{u}]_\Gamma = 0,\quad \mathbf{u} \cdot \mathbf{n} = u_\Gamma,\quad
	\left[-p\mathbf{n}+\mu\left(\nabla \mathbf{u}  +(\nabla \mathbf{u})^T\right)\mathbf{n}
	\right]_\Gamma = 
	-\sigma \kappa \mathbf{n},	
	\end{equation}
where $[\cdot]_\Gamma$ denotes the jump of the quantity across the interface, $\mathbf{n}$ and $u_\Gamma$ are the interface normal vector and velocity, $\sigma$ is the surface tension and $\kappa$ is the interface curvature. The first two conditions in (\ref{eq2_2}) describe the continuity of fluid velocity and mass conservation, while the last condition corresponds to the balance of momentum. The no-slip condition, $u=0$, is assumed at a rigid boundary. This condition is the simplest choice for the boundaries, which is also convenient from a numerical point of view. We stop our simulations before the mixing layer reaches the boundaries.

We assume the Boussinesq approximation, valid 
for small Atwood numbers $\mathcal{A} = (\rho_1-\rho_2)/(\rho_1+\rho_2) \ll 1$. It corresponds to the density 
treated as a constant and density variations affecting only the buoyancy force as
	\begin{equation}
	\label{eq2_1b}
	\rho = \rho_0,\quad \mathbf{f} = -\rho_0 \theta \tilde{g}\,\mathbf{e}_y,
	\end{equation}
where $\rho_0 = (\rho_1+\rho_2)/2$ is a background density, $\tilde{g} = \mathcal{A}g$ is the effective gravity, $\mathbf{e}_y = (0,1)$ is the unit vector in vertical direction, and $\theta$ is the order parameter equal to $1$ in the first phase and $-1$ in the second phase. In this formulation, the background value of the buoyancy term $\rho_0\mathbf{g}$ is included into the pressure. If viscosities $\mu_1$ and $\mu_2$ of two components are close, one can use the mean kinematic viscosity $\nu = (\mu_1+\mu_2)/(2\rho_0)$.  For the study of the Rayleigh-Taylor systems without the assumption of the Boussinesq approximation, we refer to the Refs.~\cite{biferale2010high,scagliarini2010lattice,goncharov2002analytical}. 

\begin{figure}[t]
\includegraphics[width=0.9\textwidth]{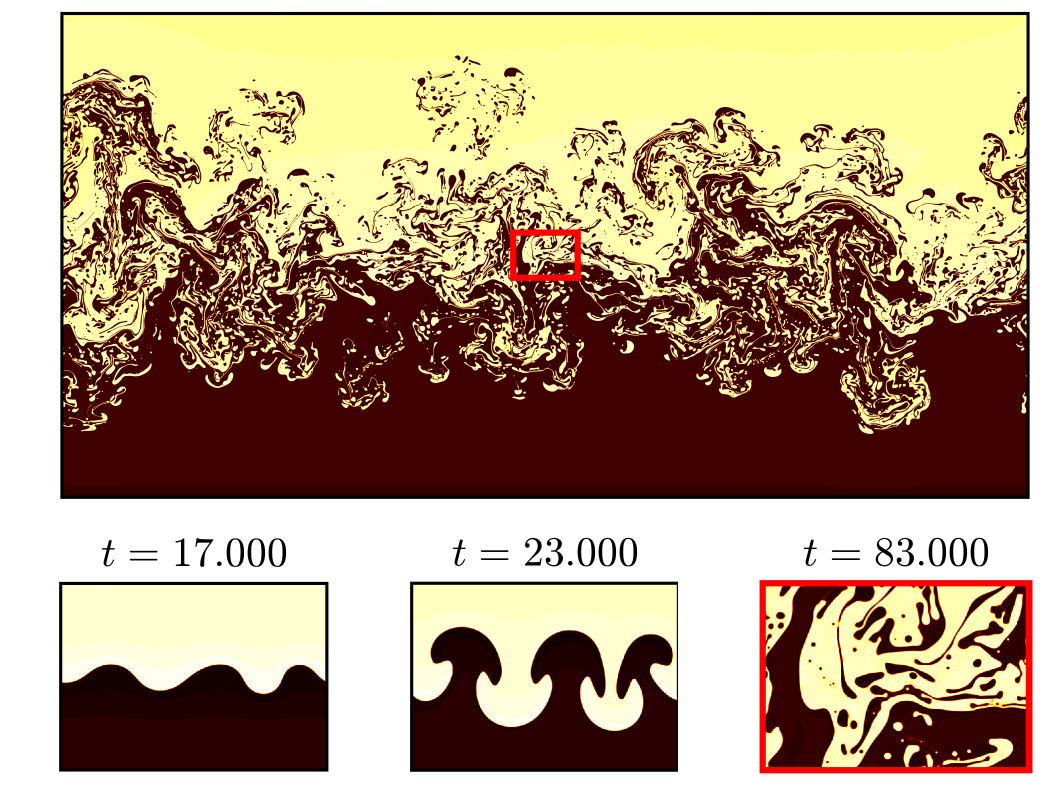}
\caption{Mixing layer of the immiscible Rayleigh-Taylor turbulence, where the yellow color represents a heavier phase and the brown color corresponds to a lighter phase. Lower pictures show the phases in the small region (marked in the center of the main panel) for three different times: the initial linear growth, formation of nonlinear mushroom-like structures at intermediate times, and fully developed turbulent mixing at larger times. Simulations are performed on the grids $10.000 
\times 5.000$ in lattice Boltzmann units (lbu), a simple artificial set of units with spatial and time steps verifying $\Delta t=\Delta x=\Delta y=1$. This set of units is directly connected with the lattice Boltzmann method described in Section~\ref{sec3}.}
\label{fig1}	
\end{figure}

Initial conditions at $t = 0$ for the Rayleigh-Taylor system correspond to the fluid at rest, $\mathbf{u} = 0$, with the heavier (first) phase occupying the upper half-plane $y > 0$ and the lighter (second) phase occupying the lower half-plane $y < 0$.
This configuration is an unstable stationary solution: small perturbations of the interface with wavenumbers  $k < \sqrt{2\rho_0\mathcal{A}g/\sigma}$ grow exponentially with a dispersion relation superiorly bounded by $\lambda(k)=-\nu k^2+\sqrt{g \mathcal{A} k-\sigma k^3/(2\rho_0)+(\nu k^2)^2}$~\cite{menikoff1977unstable,celani2006rayleigh,sohn2009effects},  see Fig.~\ref{Dispersion relation} in Section III. Depending on the values of viscosity and surface tension, this upper bound can be a good approximation of the actual dispersion relation~\cite{celani2009phase}. In Fig.~\ref{Dispersion relation}, it is also possible to see that the main effect of the viscosity is a small reduction of the growth rate of the instability.
After an initial linear growth such perturbations develop into nonlinear mushroom-like structures evolving further to the fully developed turbulent mixing layer, as shown in the Figures~\ref{fig1} and~\ref{fig1_velociity_components}.

\begin{figure}[t]
\includegraphics[width=0.9\textwidth]{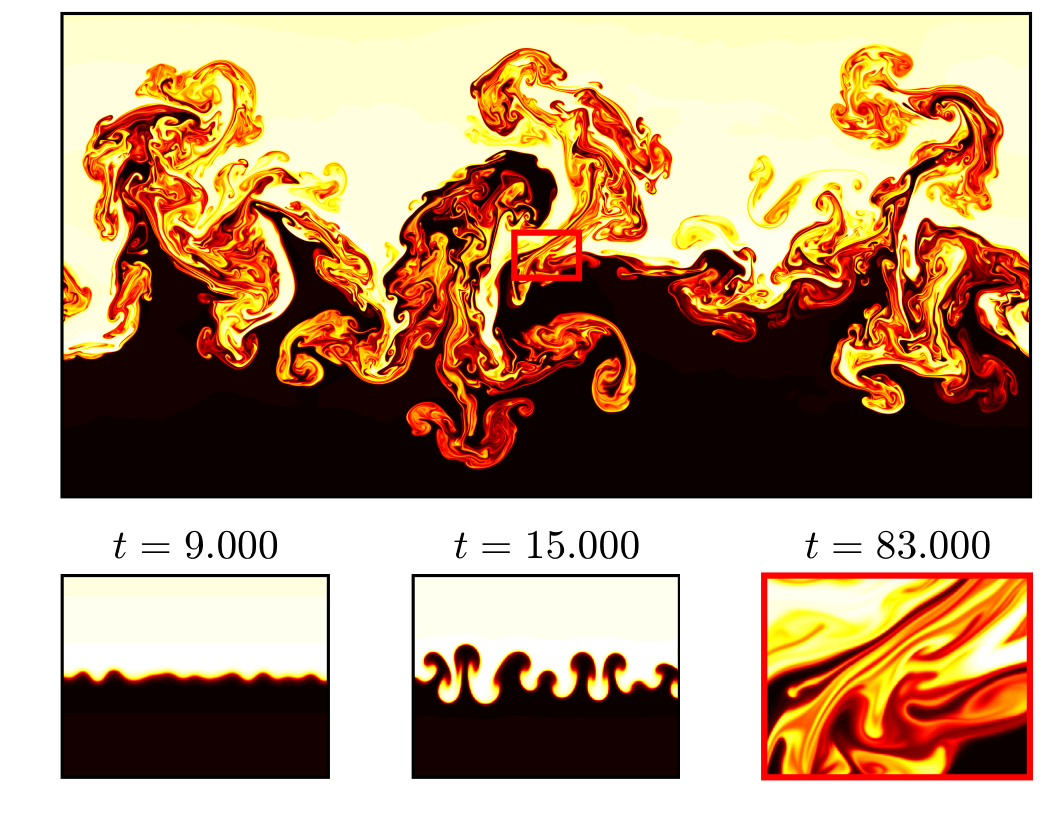}
	\caption{Mixing layer of the miscible Rayleigh-Taylor turbulence, where colors describe the fluid density; lighter colors represent a heavier fluid. Lower pictures show the densities in the small region (marked in the center of the main panel) for three different times: the initial linear growth, formation of nonlinear mushroom-like structures at intermediate times, and fully developed turbulent mixing at larger times.  Simulations are performed on the grids $10000 
\times 5000$ in lattice Boltzmann units (lbu), a simple artificial set of units with spatial and time steps verifying $\Delta t=\Delta x=\Delta y=1$. This set of units is directly connected with the lattice Boltzmann method described in Section~\ref{sec3}.}
	\label{fig2}	
\end{figure}



In the miscible flow, the fluid is modeled by a single phase with a variable density. We write this density,  also assuming the Boussinesq approximation, as $\rho = \rho_0\left(1+\mathcal{A}\theta\right)$ with the Atwood number describing a typical amplitude of density variations. The function $\theta(\mathbf{x},t)$ describing density variations satisfies the transport equation 
	\begin{equation}
	\label{eq2_4}
	\frac{\partial \theta}{\partial t}+\mathbf{u} \cdot \nabla \theta
	= \nabla \cdot (D\nabla \theta),
	\end{equation}
where $D$ is the diffusion coefficient. In general, both viscous and diffusion coefficients are functions of density. Analogous formulation arises when the density is considered to be a function of temperature $T$, in which case $\theta = -\beta (T-T_0)$ with the coefficient of thermal expansion $\beta$~\cite{landau2013fluid}. In the Boussinesq approximation, one considers a constant density and buoyancy term (\ref{eq2_1b}).

\begin{figure}[t]
	\centering
	\subfigure{\includegraphics[scale=0.36]{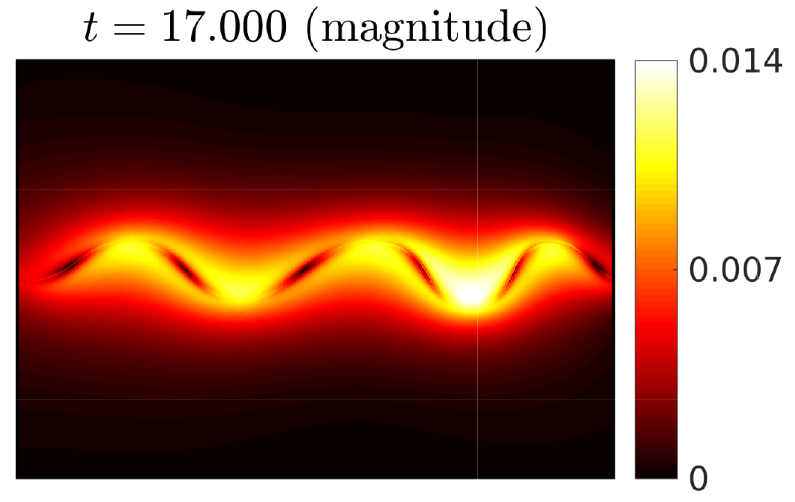}}
	\subfigure{\includegraphics[scale=0.36]{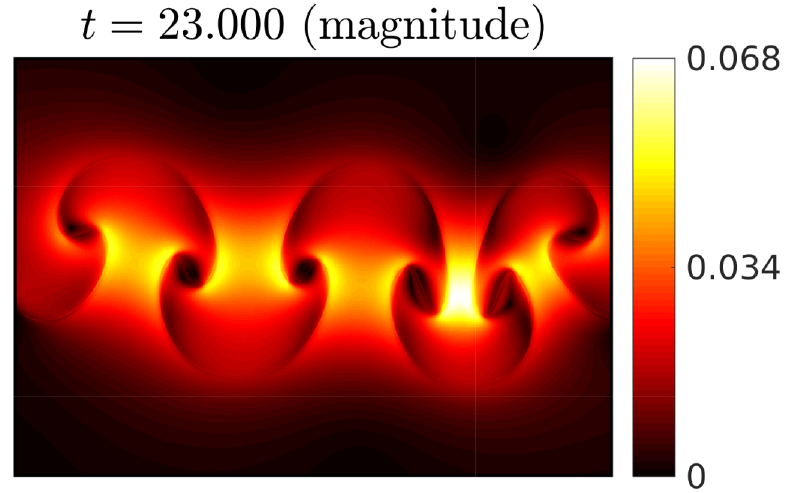}}
	\subfigure{\includegraphics[scale=0.36]{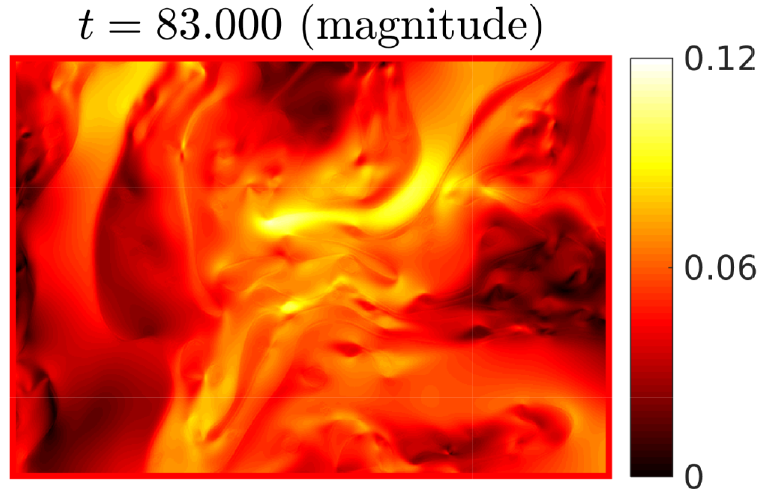}}\\
	\subfigure{\includegraphics[scale=0.36]{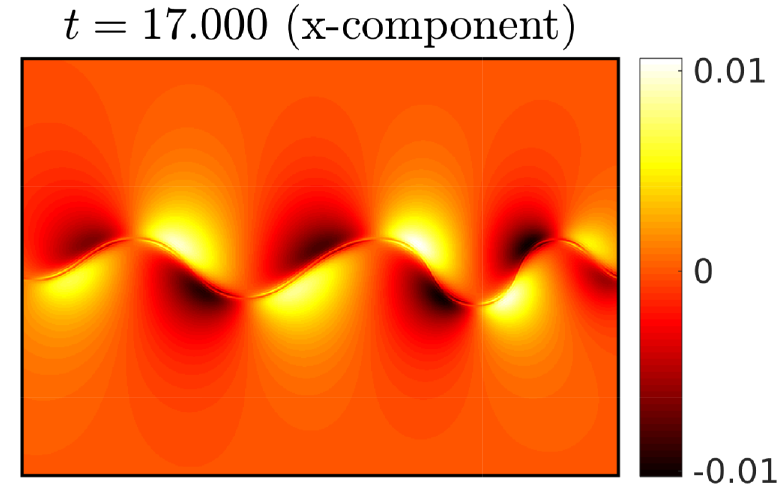}}
	\subfigure{\includegraphics[scale=0.36]{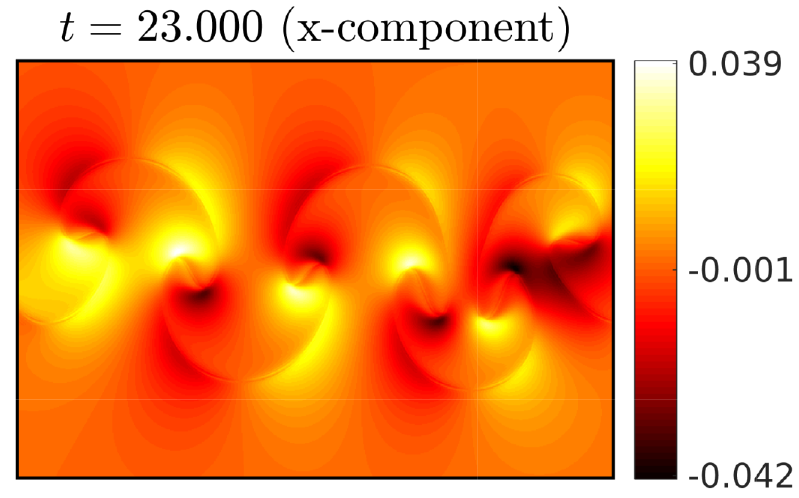}}
	\subfigure{\includegraphics[scale=0.36]{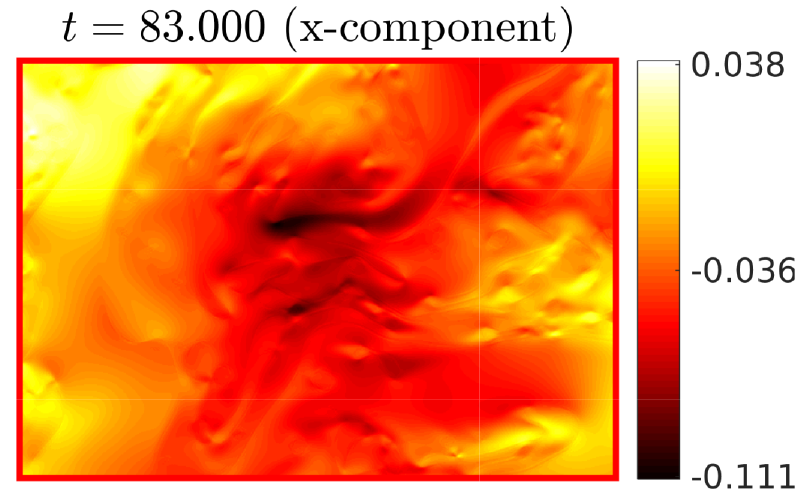}}\\
	\subfigure{\includegraphics[scale=0.36]{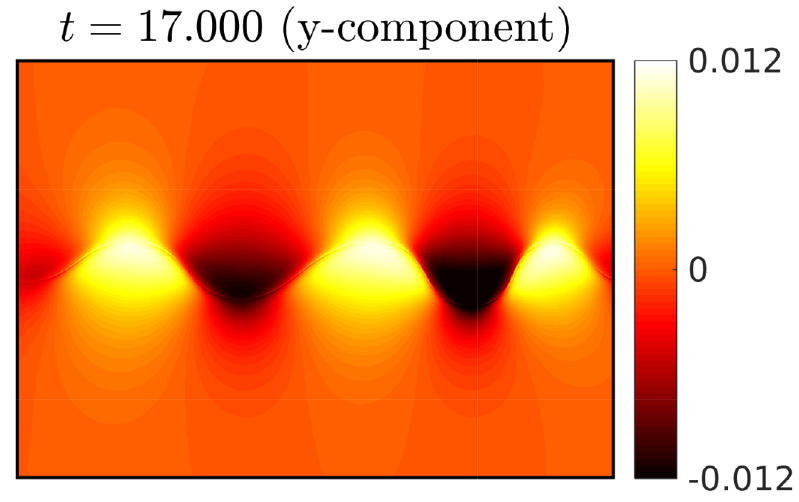}}
	\subfigure{\includegraphics[scale=0.36]{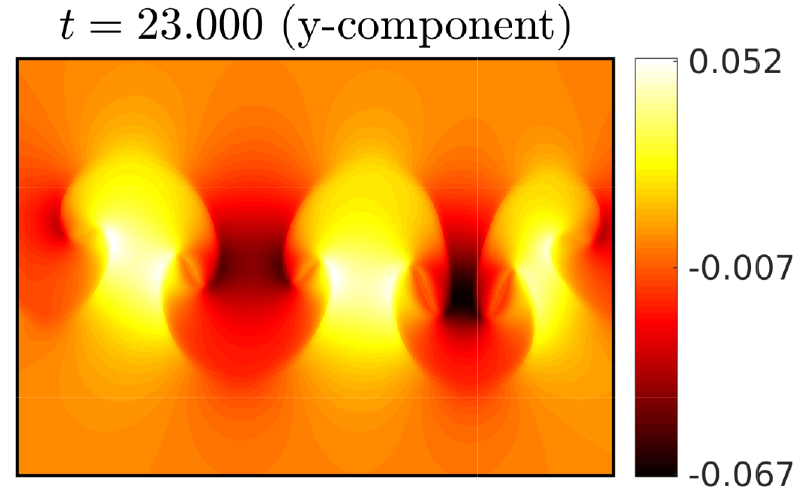}}
	\subfigure{\includegraphics[scale=0.36]{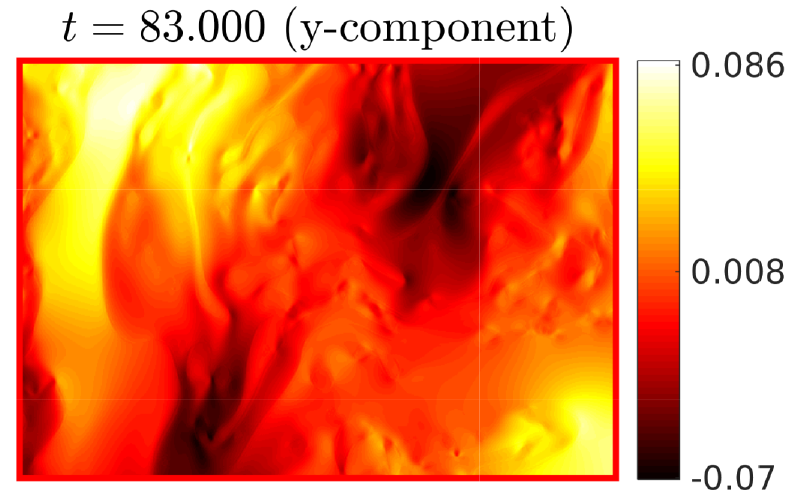}}
	\caption{Components of the velocity field for the immiscible Rayleigh-Taylor flow shown in Fig~\ref{fig1}. The velocities are indicated in simulation units.}
	\label{fig1_velociity_components}
\end{figure}

\begin{figure}[t]
	\centering
	\subfigure{\includegraphics[scale=0.36]{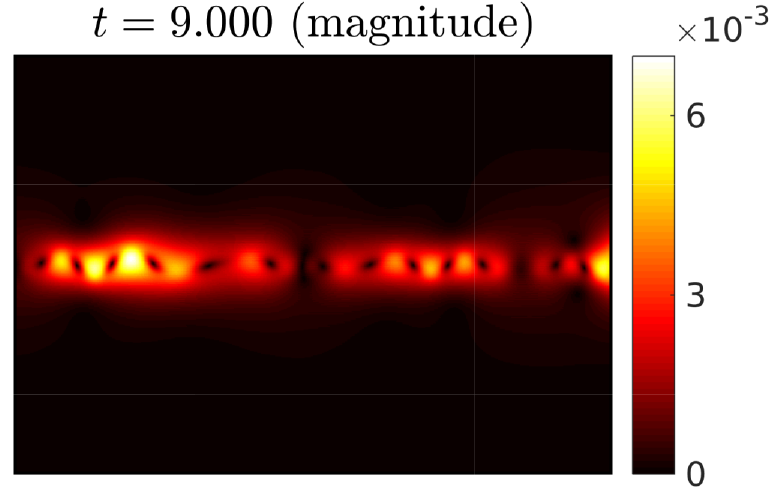}}
	\subfigure{\includegraphics[scale=0.36]{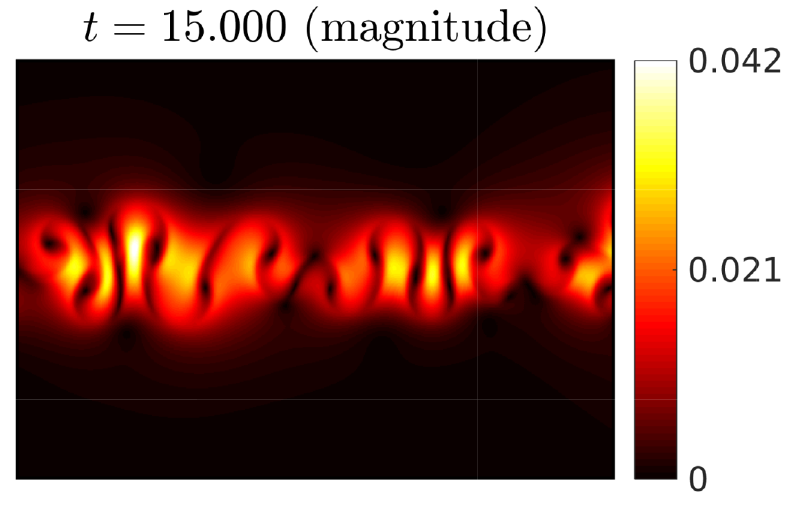}}
	\subfigure{\includegraphics[scale=0.36]{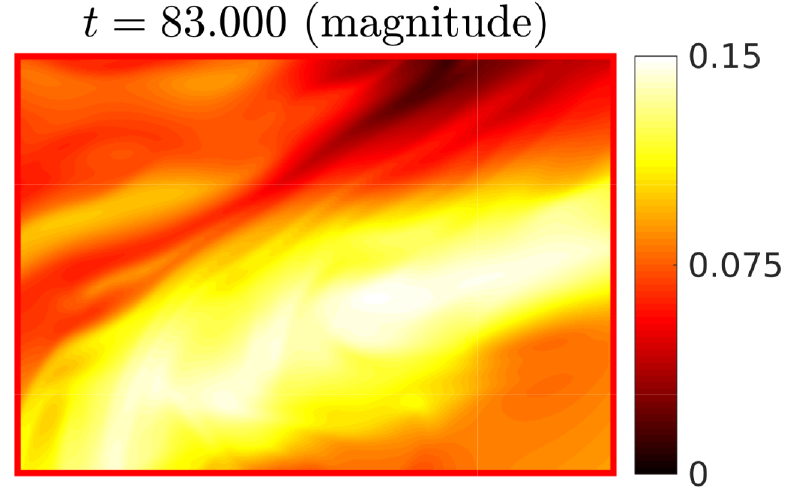}}\\
	\subfigure{\includegraphics[scale=0.36]{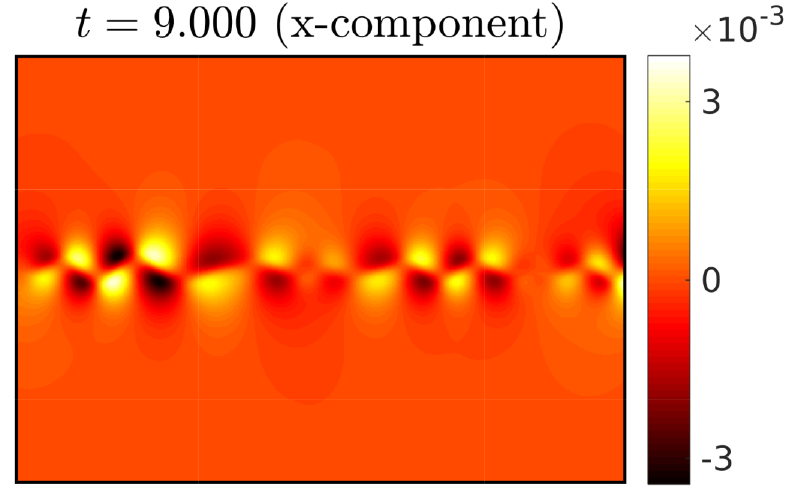}}
	\subfigure{\includegraphics[scale=0.36]{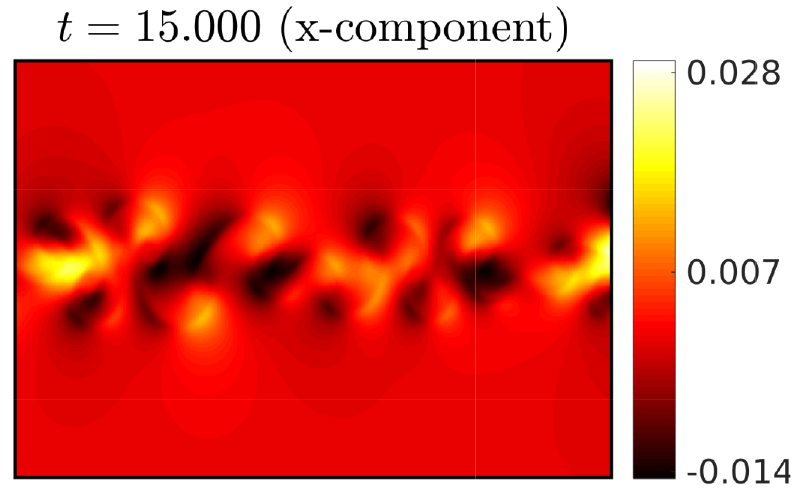}}
	\subfigure{\includegraphics[scale=0.36]{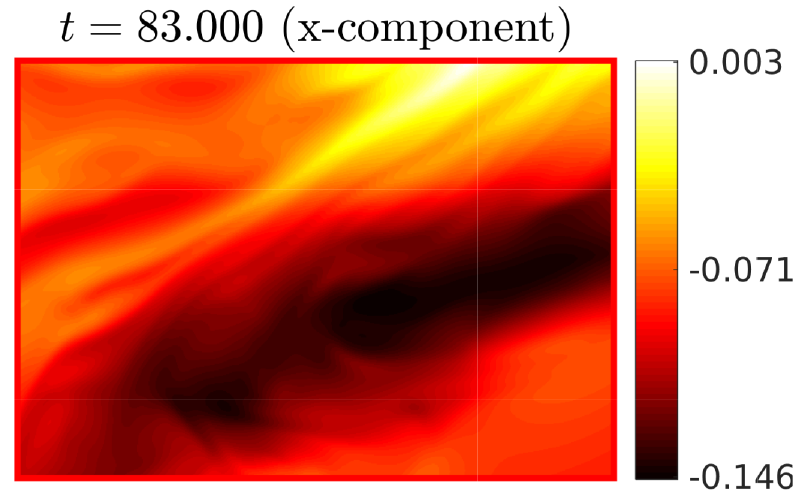}}\\
	\subfigure{\includegraphics[scale=0.36]{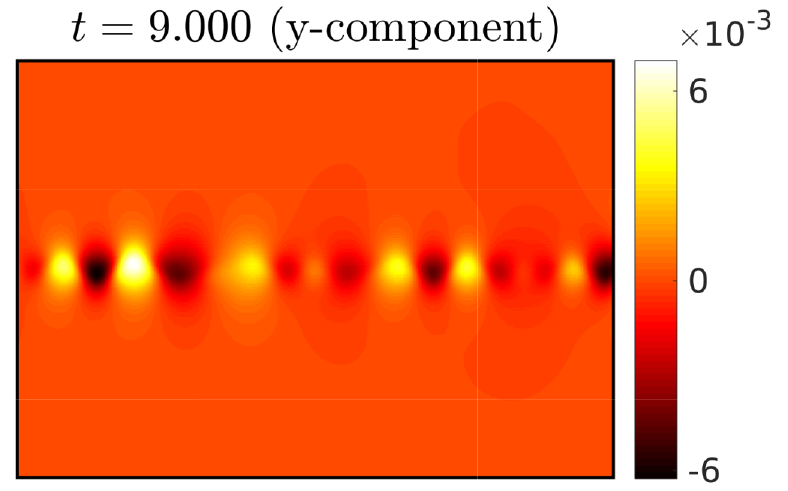}}
	\subfigure{\includegraphics[scale=0.36]{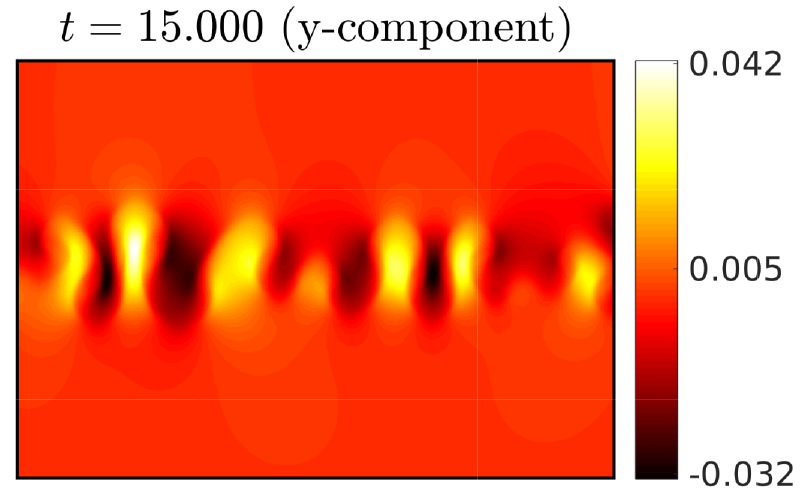}}
	\subfigure{\includegraphics[scale=0.36]{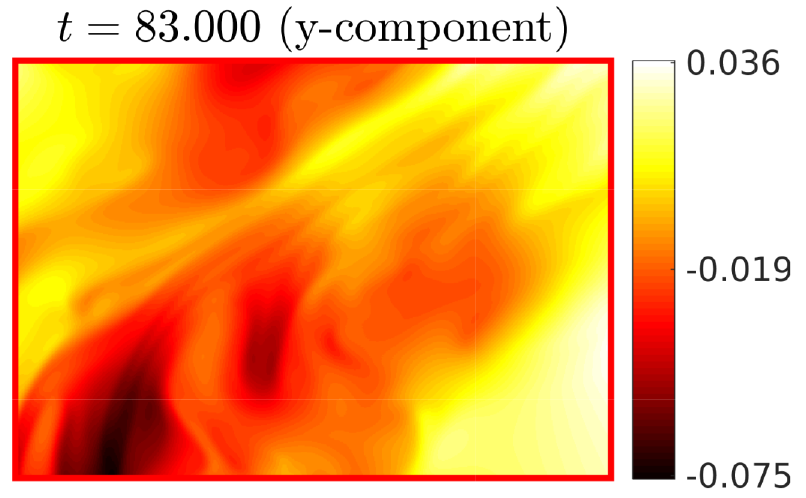}}
	\caption{Components of the velocity field for the miscible Raleigh-Taylor flow shown in Fig.~\ref{fig2}. The velocities are indicated in simulation units.}
	\label{fig_2_velocity_components}
\end{figure}

The miscible Rayleigh-Taylor instability corresponds to the same initial conditions as the immiscible one. It follows a similar scenario, where small perturbations of the interface are amplified first linearly and then nonlinearly, growing into the developed turbulent mixing layer, as shown in the Figures~\ref{fig2} and~\ref{fig_2_velocity_components}. The important difference between the immiscible and miscible cases can be seen at small scales. The immiscible Rayleigh-Taylor turbulence leads to the formation of an emulsionlike state with a multitude of small bubbles. The miscible Rayleigh-Taylor turbulence develops sharp gradients leading the enhanced diffusion at small scales.

\section{Lattice Boltzmann model}
\label{sec3}

In this section, we describe the two-component lattice Boltzmann method for simulating immiscible and miscible Rayleigh-Taylor systems in the Boussinesq approximation; we refer the reader to the Refs.~\cite{kruger2017lattice,succi2018lattice} for more details. In this method, spatial coordinates and time take values on the lattice with spacings $\Delta x$ and $\Delta t$, and the system is described by the interactions between two species of particles, A and B. Considering the so-called D2Q9 scheme, each particle is allowed to have nine velocities $\mathbf{c}_0,\ldots,\mathbf{c}_8$. These velocities are given by the vectors $(0,0)$, $(\pm c,0)$, $(0,\pm c)$ and $(\pm c,\pm c)$ with $c = \Delta x/\Delta t$, such that a particle either stays at the same lattice point or moves to a neighboring lattice point in a single time step. The system is described by the functions $f^s_i(\mathbf{x},t)$ determining the number of particles of component $s = A$ or $B$ and velocity $\mathbf{c}_i$ at a given point and time. The densities of each component and common velocity of the fluid are defined as
	\begin{equation}
	\rho_s(\mathbf{x},t)=\sum_{i} f_{is}(\mathbf{x},t),  \quad 
	\mathbf{u}(\mathbf{x},t) 
	=\dfrac{\sum_{s,i} f_i^s(\mathbf{x},t)\mathbf{c}_i/\tau_s}{\sum_{s}
	\rho_s(\mathbf{x},t)/\tau_s},
	\end{equation}
where $s = A, B$ and $i = 0,\ldots,8$. The total density is given by the sum $\rho = \rho_A+\rho_B$.

The evolution is governed by the lattice-Boltzmann equations with the Bhatnagar-Gross-Krook collision term~\cite{scarbolo2013unified} 
	\begin{equation}\label{Kinetic equation}
	f^s_i(\mathbf{x}+\mathbf{c}_i\Delta t,t+\Delta t)-f^s_i(\mathbf{x},t)
	=-\dfrac{1}{\tau_s}\left[ f^s_i(\mathbf{x},t)
	-f_i^{s(eq)}(\rho_s,\mathbf{u}+\tau_s\mathbf{F}_s/\rho_s )\right], 
	\end{equation}  
where $\tau_s$ and $\mathbf{F}_s$ are the relaxation time and the forcing term for component $s$, respectively. 
The right-hand side in (\ref{Kinetic equation}) describes the relaxation towards the local equilibrium distribution 
	\begin{equation}
	\label{Equilibrium distribution}
	f_i^{s(eq)}(\rho_s,\mathbf{u}')
	=\rho_s w_i\left(1+\dfrac{3\mathbf{c}_i\cdot \mathbf{u}'}{c^2}
	+\dfrac{9(\mathbf{c}_i\cdot \mathbf{u}')^2}{2 c^4}
	-\dfrac{3\mathbf{u}' \cdot \mathbf{u}'}{2c^2} \right),\quad
	\mathbf{u}' = \mathbf{u}+\frac{\tau_s\mathbf{F}_s}{\rho_s},
	\end{equation}
with the lattice sound speed $c_s =c/\sqrt{3}$ and constant weights $w_i$. These weights are expressed through velocity components $\mathbf{c}_i = (c_i^1,c_i^2)$ by the conditions
	\begin{equation}
	\sum_{i}w_ic_i^ac_i^b = c_s^2\delta_{ab},\quad
	\sum_{i}w_ic_i^ac_i^bc_i^cc^d_i = c_s^4\left(\delta_{ab}\delta_{cd}
	+\delta_{ad}\delta_{bc}+\delta_{ac}\delta_{bd}\right)
	\quad
	\textrm{for}\ \ a,b,c,d = 1,2,
	\end{equation}
where $\delta_{ab}$ is the Kronecker delta.

The forcing terms $\mathbf{F}_s = \mathbf{F}_{s}^{\textrm{ff}}+\mathbf{F}_{s}^{\textrm{fb}}+\mathbf{F}_{s}^{\textrm{ext}}$
contain three parts describing the fluid-fluid interaction, the fluid-boundary interaction and the external forces. The first is given by the Shan-Chen inter-molecular force as
	\begin{equation}\label{Shan-Chen forcinng term}
	\mathbf{F}_{s}^{\textrm{ff}}(\mathbf{x},t)
	= - G_{AB}\rho_s(\mathbf{x},t)
	\sum_{i}w_i\rho_{s'}(\mathbf{x}+\mathbf{c}_i\Delta t,t)\mathbf{c}_i,
	\end{equation}
with $s' = B$ and $s = A$ or vice versa. Here, we consider a system without self-interaction, where the coupling constant $G_{AB}$ controls the interaction between components $A$ and $B$. 
The interaction between fluid and boundary is given by
	\begin{equation}
	\mathbf{F}_s^{\textrm{fb}}
	=-G_{sb}\rho_s(\mathbf{x},t)\sum_{i}w_iS(\mathbf{x}+\mathbf{c}_i\Delta t)\mathbf{c}_i,
	\end{equation}
where $S(\mathbf{x})$ is the indicator equal to unity at boundary nodes and vanishing otherwise. The parameters $G_{Ab}$ and $G_{Bb}$ control interactions between fluid components and solid boundary; they relate to contact angles of fluids in the mixture. External forces are introduced as 
	\begin{equation}
	\label{eq3_Fex}
	\mathbf{F}^{\textrm{ext}}_A= -\rho_A \tilde{g} \,\mathbf{e}_y,
	\quad
	\mathbf{F}^{\textrm{ext}}_B= \rho_B \tilde{g}\,\mathbf{e}_y,
	\end{equation}
which yield the buoyancy forces in Boussinesq approximation, as we will see below.

\subsection{Implementation details}

We choose $\Delta x = \Delta t =1$ (considered as lattice-Boltzmann units) in the rectangular
domain of horizontal size $L_x = 10^4$ and vertical size $L_y = L_x/2$. Periodic boundary conditions are assumed in the horizontal direction with the solid bottom and top boundaries. The bounce-back relation \cite{succi2018lattice,li2020multiscale} is used for the distribution function $f^s_i(\mathbf{x},t)$ at the solid boundaries for modeling the no-slip condition. The  relaxation time $\tau = 0.53$ is chosen for both components, providing the kinetic viscosity $\nu = c_s^2(\tau-1/2) = 0.01$. In the continuous limit, the lattice Boltzmann system approximates the coupled Navier-Stokes and Cahn--Hillard equations~\cite{succi2018lattice,benzi2009mesoscopic} for the velocity field $\mathbf{u}(\mathbf{x},t)$, the total density $\rho(\mathbf{x},t)$ and the order parameter $\phi(\mathbf{x},t) = \rho_A-\rho_B$. For small fluid velocities (small lattice Mach numbers) $|\mathbf{u}|\ll c_s$, the flow can be assumed incompressible. We consider pure densities of both fluid components equal to $1.10$ and the gravity parameter $\tilde{g} = 9\cdot 10^{-6}$. Since changes of the total density due to pressure variations and mixing are small, we approximate $\rho(\mathbf{x},t) \approx \rho_0$ by a constant. In this case, the Boussinesq buoyancy force (\ref{eq2_1b}) agrees with our choice of the external force (\ref{eq3_Fex}) for $\theta = \phi/\rho_0$. 

The coupling constant $G_{AB}$ has a critical value with the immiscible (two phase) fluid for stronger couplings and miscible (single phase) fluid for weaker couplings. For our immiscible and miscible models, we select $G_{AB}=0.1381$ and $G_{AB}=0.0805$, respectively.  In the interactions with the boundaries, we use neutral wetting, i.e., $G_{Ab}=G_{Bb}=0$, to minimize the influence of the boundaries in the simulations. In the immiscible model, two phases are separated by a diffuse interface having a width of approximately $l_{int} \sim 3$ grid nodes. This model approximates the Boussinesq system (\ref{eq2_1})--(\ref{eq2_1b}) considered at scales much larger than $l_{int}$ with the surface tension $\sigma = 0.0059$ obtained from pressure measurements for large bubbles. Similarly, one recovers the miscible Boussinesq system (\ref{eq2_1}), (\ref{eq2_1b}) and (\ref{eq2_4}) in the continuous limit for small gradients of the order parameter. The diffusion coefficient can be estimated roughly as $D \simeq c_s^2\left[ \left(\tau -1/2\right) -\rho  \tau G_{AB}/2 \right] = 0.002$~\cite{benzi2009mesoscopic}. Though the diffusion coefficient is a function of the order parameter in a more accurate description, such dependence is not important for our study based on the phenomenological theory of turbulence.

Simulations are implemented on GPUs of the model NVIDIA Tesla V100 PCIe 32 GB. The use of a GPU is instrumental to accumulate better statistics with a reasonable amount of time.
Specifically, for our main tests we consider ensembles with 15 simulations on the grids $10000 
\times 5000$ for the immiscible and miscible  flows performed for different random initial disturbances. For further quantitative indications on the performances and potentialities of the GPU codes, we refer the reader to Refs.~\cite{bernaschi2009graphics,bernaschi2017gpu,pelusi2019impact}. The choice of the size of the ensembles is motivated by  small values of standard deviations verified in our numerical experiments, indicating a small dependence on the initial conditions for big computational grids like the ones used by us. For smaller grids and early stages of turbulence, the influence of initial conditions was studied in~\cite{meshkov2019group,biferale2018rayleigh}.

\begin{figure}
     \centering
     \includegraphics[scale=0.7]{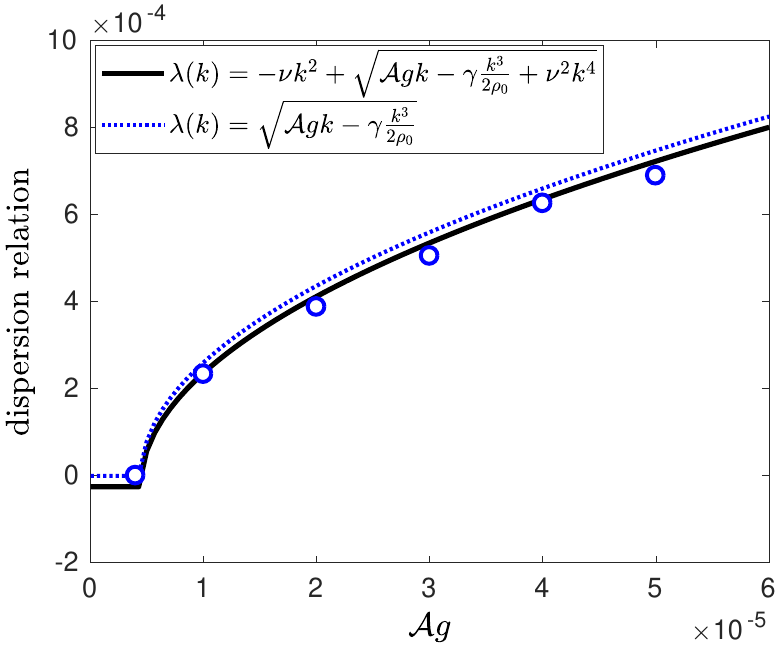}
     \caption{  We show the theoretical upper bound for the dispersion relation $\lambda = -\nu k^2+\sqrt{g \mathcal{A} k-\sigma k^3/(2\rho_0)+(\nu k^2)^2}$~\cite{menikoff1977unstable} for the immiscible RT system (black curve) compared to the results of lattice Boltzmann simulations (circles) obtained by measuring exponential growth of the maximum interface displacement for different values of the effective gravity $\tilde{g} = \mathcal{A}g$. Error bars for fitting are approximately the size of the data symbols. We also compare with the classical dispersion relation without viscosity~\cite{celani2009phase}, indicated by the blue dashed line, showing that in our case the main effect of the viscosity is a small reduction in the growth rate of the RT instability.  The simulations are performed on grids of size $512 \times 512$ with parameters corresponding to the relaxation time $\tau=1.0$ and interaction parameter $G_{AB}=1.22$, which gives the kinematic viscosity $\nu=0.1667$ and the surface tension coefficient  $\sigma=0.061$. The numerical experiments considered different values of $\tilde{g}$ for a fixed $k=2\pi/512$. }
     \label{Dispersion relation}
 \end{figure}

We perform a number of additional numerical tests justifying the validity of the lattice Boltzmann model for the Rayleigh-Taylor instability. In particular, we show that numerical dispersion relations of the initial linear instability are in agreement with theoretical predictions~\cite{celani2009phase,sohn2009effects}; see Fig.~\ref{Dispersion relation}. We verify that 
 non-isotropic contributions to the stress tensor due to variations of the order parameter are small in the miscible case. In the immiscible flow, these contributions grow in time following the increase of the interface, but they remain small compared to buoyancy and viscous contributions. Also, numerical anisotropy of the Shan-Chen force generates spurious currents~\cite{sbragaglia2007generalized,connington2012review} within thin diffuse interfaces, which do not affect most of our measurements but may interfere in the results for enstrophy, as discussed in the end of Section~\ref{sec_IC}. A more detailed account of the tests describing the validity and performance of the numerical method will be given elsewhere. For simulations in this paper, we initialize the flow by using an equilibrium immiscible configuration and adding a small random (white-noise) deformation to the interface with an amplitude of 4 grid points. In this equilibrium configuration, the first phase consists primarily of component $A$ with about $9\%$ of component $B$, and vise versa for the second phase.
 
\section{Evolution and shape of the mixing layer}

In this section, we investigate the large-scale dynamics of the RT mixing layer, comparing its development in immiscible and miscible flows. 

The development of the mixing layer from a small initial perturbation of the straight interface line is presented in Fig.~\ref{fig1} (immiscible) and Fig.~\ref{fig2} (miscible). The panels in the bottom of these figures correspond to zooms of a small region in the middle of the computational domain (red rectangles in the main plots) at different times. They illustrate the initial linear growth of perturbations, which develop into a nonlinear quasi-periodic pattern with mushroom-like structures. For later times, these structures break down, forming a fully developed  turbulent mixing layer. 

The macroscopic properties of the turbulent mixing layer are described by its width $L(t)$ and the large-scale velocity fluctuation $U(t)$. The latter estimates the velocity of large-scale plumes within the mixing layer, which yields the relation $U(t) \sim dL/dt$. Phenomenologically, the energy balance $dE/dt \sim -dP/dt$ describes the transfer of potential energy $P \propto -\mathcal{A} g L$ into kinetic energy $E \propto U^2$; see e.g., Ref.~\cite{boffetta2017incompressible}. Recall that the Atwood number $\mathcal{A}$ characterizes typical density variations, and we denoted $\tilde{g} = \mathcal{A} g$ in the Boussinesq approximation and the lattice Boltzmann method. The energy balance provides the relation $dU/dt \sim \mathcal{A} g$. Integrating, we obtain the quadratic asymptotic growth of the mixing layer and linear growth of the velocity fluctuation as
	\begin{equation}
	\label{eq4_1}
	L(t) \approx \alpha_L \mathcal{A} g t^2,\quad
	U(t) \approx \alpha_U \mathcal{A} g t,
	\end{equation} 
where the starting moment is set to $t = 0$. The two dimensionless parameters $\alpha_L$ and $\alpha_U$ characterize the efficiency of the conversion of potential into kinetic energy. 

\begin{figure}
\includegraphics[width=0.55\textwidth]{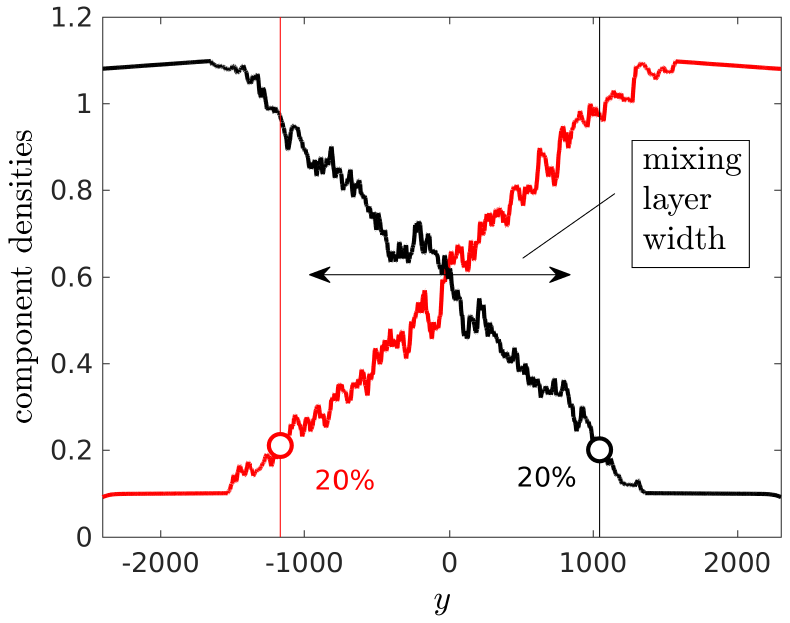}
\caption{Definition of the mixing layer as the region between two points, where the averaged component densities $\rho_A$ (red) and $\rho_B$ (black) attain $20\%$ of the total density.}
\label{Nixing_layer_def}
\end{figure}

The numerical procedure for the analysis of the mixing layer is illustrated in Fig.~\ref{Nixing_layer_def}. Here the red and black lines show the dependence on the vertical coordinate $y$ for the  component densities $\rho_A(\mathbf{x},t)$ and $\rho_B(\mathbf{x},t)$ averaged with respect to the horizontal coordinate $x$. We define the mixing layer as the region between two points, at which the averaged density of each component reaches 20\% of the total density. This definition separates the central region of the mixing layer, cutting off its most non-homogeneous outer parts. Then, the large-scale velocity fluctuation is introduced as $U^2 = \langle \|\mathbf{u}\|^2 \rangle_{\textrm{ML}}$, where the averaging is performed within the central region of the mixing layer.

\begin{figure}
\includegraphics[width=0.39\textwidth]{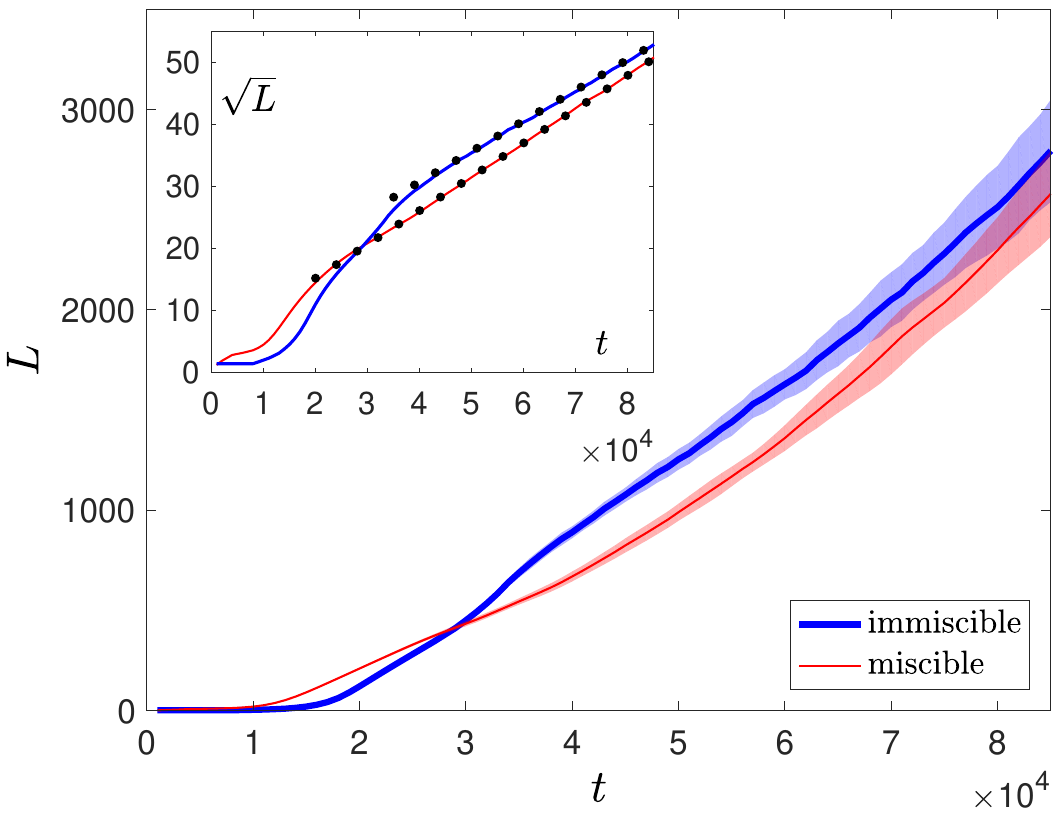}
\put(-90,148){(a)}
\put(130,148){(b)}
\hspace{0.9cm}
\includegraphics[width=0.4\textwidth]{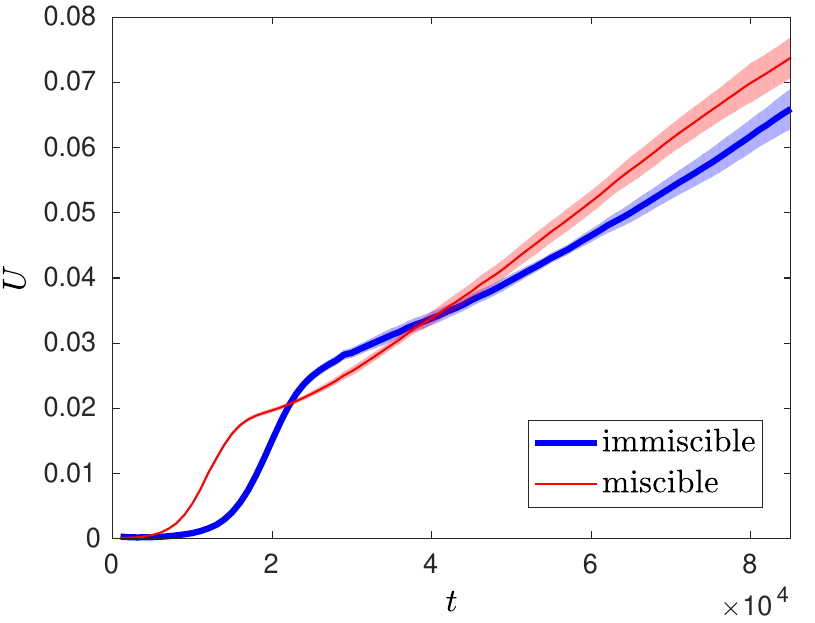}
\caption{(a) Width of the mixing layer $L(t)$ and (b) large-scale velocity fluctuation $U(t)$ depending on time for immiscible (bold blue) and miscible (thin red) flows. Shaded areas indicate standard deviations. The inset in figure (a) compares the graphs $\sqrt{L(t)}$ in the region of turbulent mixing with the estimated slopes (\ref{Parameter alpha}) shown by dotted lines.}
\label{fig4}
\end{figure}

Numerical measurements for the width $L(t)$ and speed $U(t)$ of the mixing layer, averaged with respect to ensembles of realizations, are presented in Fig.~\ref{fig4} for both immiscible and miscible flows.  We associate the beginning of turbulent mixing with the time when mushroom-like structures break down into a chaotic multi-scale mixing layer; see Figs.~\ref{fig1} and \ref{fig2}. In our simulations, turbulent mixing layers develop roughly at the times $t \gtrsim 4\times 10^4$ in the immiscible case and $t \gtrsim 3\times 10^4$ in the miscible case. The difference between these initial times can be attributed to the resistance caused by the surface tension in immiscible flows. 
 All simulations are stopped at times $t \approx 8.5\times 10^4$. For larger times, the mixing layer may be affected considerably by the top and bottom rigid boundaries. In terms of the Reynolds number $\mathrm{Re} = UL/\nu$, the developed turbulent regime corresponds to $(0.3 \sim 2.1) \times 10^4$ for the immiscible flow and $(0.1 \sim 2.1) \times 10^4$ for the miscible flow.

\begin{figure}
\includegraphics[width=0.4\textwidth]{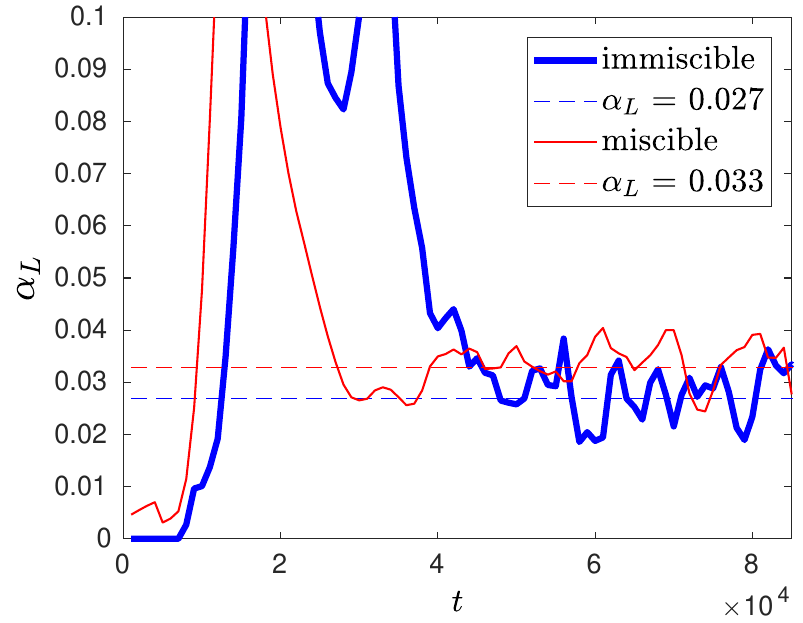}
\put(-90,149){(a)}
\put(125,149){(b)}
\hspace{0.8cm}
\includegraphics[width=0.4\textwidth]{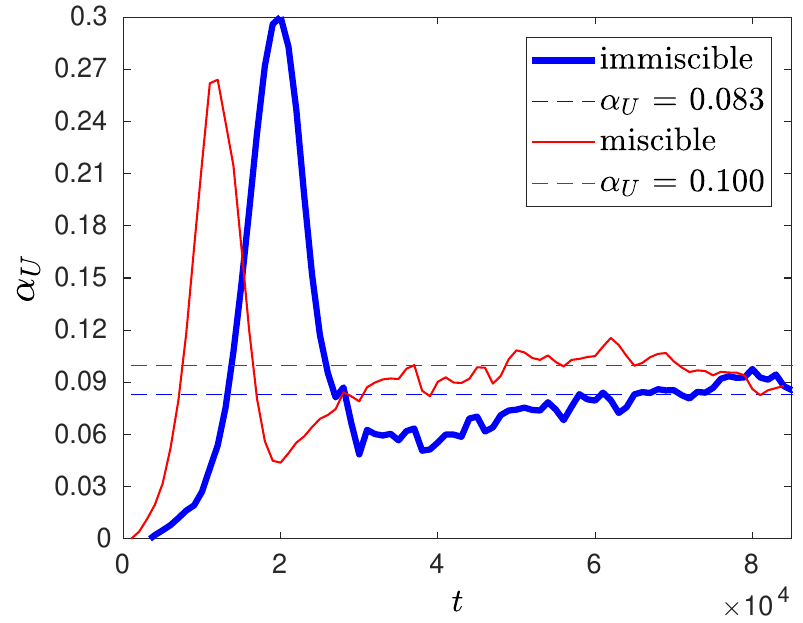} 
\caption{Measurement of the dimensionless pre-factors for the immiscible (bold blue) and miscible (thin red) flows: (a) $\alpha_L$ for the mixing layer width and (b) $\alpha_U$ for the large-scale velocity fluctuation. Constant values (dashed lines) are estimated in the regions of turbulent mixing.}
\label{fig5}
\end{figure}

In order to verify the phenomenological predictions (\ref{eq4_1}), we estimate
	\begin{equation}\label{Parameter alpha}
	\alpha_L = \frac{1}{4\mathcal{A}gL}\left(\frac{dL}{dt}\right)^2, \quad
	\alpha_U = \frac{1}{\mathcal{A}g} \frac{dU}{dt},
	\end{equation}
where the derivatives are computed by finite differences.
Such relations are more robust numerically because they are insensitive to shifts of the initial time, $t \mapsto t-t_*$, accounting for the early non-turbulent development of the mixing layer. The results of computations with formulas (\ref{Parameter alpha}) are shown in Fig.~\ref{fig5}, demonstrating clear tendencies to constant values in the regions of developed turbulent mixing. The estimated values are $\alpha_L = 0.027 \pm 0.005$ and $\alpha_U = 0.083 \pm 0.007$ for immiscible and $\alpha_L = 0.033 \pm 0.004$ and $\alpha_U = 0.1 \pm 0.005$ for miscible flows; see also the direct comparison in the inset of Fig. \ref{fig4}(a). Notice that previous experiments~\cite{clark2003numerical,scagliarini2010lattice,celani2006rayleigh,boffetta2017incompressible}  reported the pre-factors $\alpha_L$ between 0.01 and 0.06 for the miscible mixing layer, which are compatible with our estimates taking into account that we use a different definition of $L$. Our results provide a value of $\alpha_L$ in the immiscible case slightly lower than those in the miscible situation, see Fig. \ref{fig5}(a), indicating that the immiscible RT turbulence may be less efficient in the conversion of potential into kinetic energy; the same conclusions are valid for the other pre-factor $\alpha_U$. However, the differences are small (comparable to standard deviations), which does not exclude the possibility that they are actually equal for immiscible and miscible flows in the asymptotic limit of an infinitely large domain. Analogous universality of the mixing layer pre-factors with respect to small-scale physics was observed recently for the Kelvin--Helmholtz instability~\cite{thalabard2020butterfly}, where Navier--Stokes flows were compared to a point-vortex model. 

\begin{figure}
\includegraphics[width=0.4\textwidth]{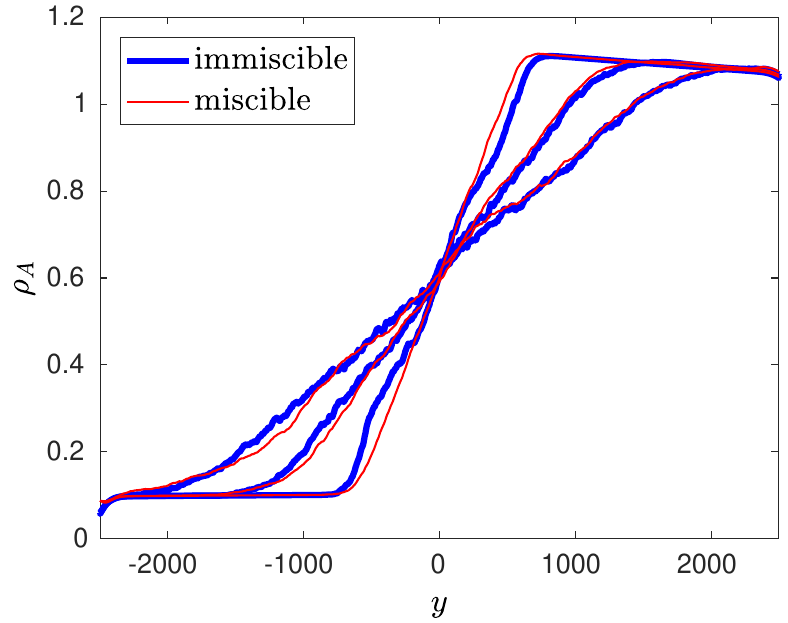}
\put(-90,150){(a)}
\put(125,150){(b)}
\hspace{0.9cm}
\includegraphics[width=0.4\textwidth]{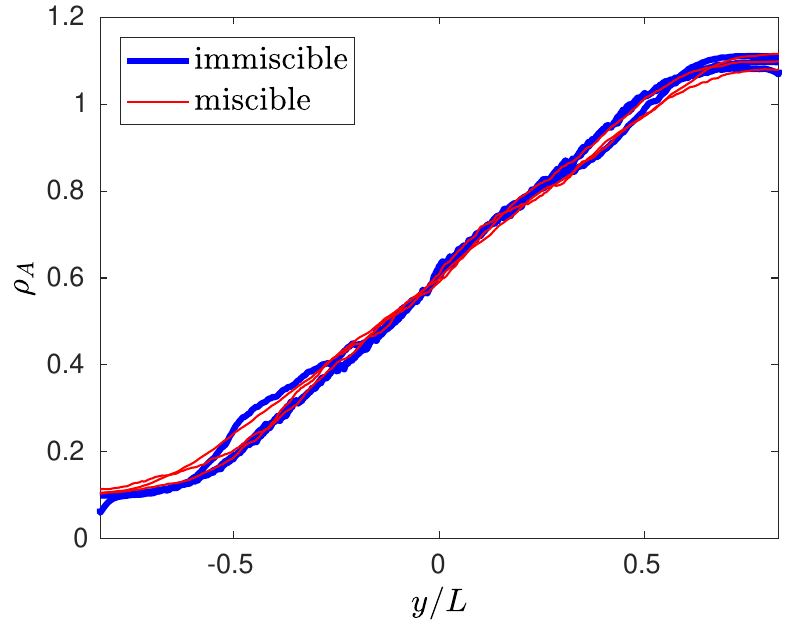} 
\caption{Density profiles for the component $A$ averaged with respect to horizontal coordinate $x$ and ensemble of realizations. The results are shown at three consecutive times $t = 4.5\times10^4$, $7\times10^4$ and $8.9\times10^4$. (a) Dependence on the vertical coordinate $y$. (b) Dependence on the rescaled vertical coordinate $y/L(t)$ demonstrates self-similarity and universality of the density profile for immiscible and miscible flows.}
\label{Average density profiles}
\end{figure}

Figure~\ref{Average density profiles} shows profiles for the density $\rho_A$ of component $A$ averaged with respect to the horizontal coordinate $x$ and an ensemble of realizations. The figure (a) shows profiles at three consecutive times both for immiscible (bold blue) and miscible (thin red) flows. By the dimensional argument leading to power laws (\ref{eq4_1}), one can also conjecture that the averaged density profiles are self-similar in the regime of developed turbulent mixing, with the dependence only on the ratio $y/L(t)$. This conjecture is supported by Fig.~\ref{Average density profiles}(b), where the graphs from the left panel collapse into a single curve when plotted with respect to the rescaled coordinate $y/L(t)$.  The graphs suggest 
that the inner region of the mixing layer develops a linear average density profile 
with a slope decreasing proportionally to $1/L(t) \propto t^{-2}$. This linear 
 profile implies statistical homogeneity inside the mixing layer~\cite{boffetta2017incompressible,celani2006rayleigh}. Notice that, up to numerical fluctuations, the self-similar profiles are indistinguishable for the immiscible and miscible cases. This provides further evidence for the universality of large-scale properties in the RT turbulence for immiscible and miscible flows.
 
  Self-similarity, homogeneity and isotropy in the statistical sense~\cite{frisch1995turbulence} are important assumptions for phenomenological theories derived similarly to the Kolmogorov's theory of turbulence (K41)~\cite{kolmogorov1991dissipation}. For miscible Rayleigh-Taylor systems, the tendency toward isotropy restoration of small-scale fluctuations has been numerically verified by the Refs.~\cite{biferale2010high,boffetta2009kolmogorov,boffetta2010statistics} and experimentally by the Ref.~\cite{ramaprabhu2004experimental}. The similarities of the statistics between miscible and immiscible  RT flows in our experiments indicate that the same tendency may also happen for the immiscible Rayleigh-Taylor systems, which motivates the definition of turbulence for the observed late-time behavior. Notice that, though numerical simulations confirm self-similar RT dynamics, some experiments report on departures from the canonical turbulence scenario with strong sensitivity to initial conditions; see e.g. \cite{meshkov2019group,meshkov2013some,robey2003time}.

\section{Evolution of the Interface in the immiscible RT turbulence}
\label{sec_IC}

An intricate evolution of the interface between two phases is the most distinctive feature of immiscible RT turbulence. In this section, we study the statistical properties of the interface depending on time and scale, the distribution of drops with respect to their size, and the effects of the interface on the flow. 

The interface evolution with the formation of drop-rich (emulsion) regions is driven by the velocity fluctuations at small scales. In the RT turbulence, such fluctuations can be described phenomenologically assuming that the dynamics at small scales adjusts in a quasi-stationary (adiabatic) manner to the large-scale growth of the mixing layer described by the width $L(t)$ and velocity $U(t)$. In two-dimensional flows, statistics at small-scales follows the so-called Bolgiano--Obukhov scenario~\cite{bolgiano1959turbulent,obukhov1959influence,siggia1994high}, which assumes the balance of buoyancy and nonlinear terms with density fluctuations cascading toward small scales at a constant rate. For equations (\ref{eq2_1})--(\ref{eq2_1b}), this balance reads $(\delta_r u)^2/r \sim \mathcal{A}g\delta_r\theta$, where we denoted coarse-grained velocity fluctuations at scales $r$ by $\delta_r u$ and analogous fluctuations of the order parameter by $\delta_r\theta$. With the estimate $\varepsilon_{\theta} \sim (\delta_r \theta)^2 (\delta_r u)/r$ for the flux of order-parameter fluctuations, elementary derivation yields the well-known Bolgiano--Obukhov scaling laws $\delta_r u \propto r^{3/5}$ and $\delta_r \theta \propto r^{1/5}$. These laws are valid at scales of the inertial interval $\eta \ll r \ll L$ limited from below by the viscous (Kolmogorov) scale $\eta$, at which viscous forces must be taken into account. There is also a limitation caused by the interface introducing the scale $\ell$ of a typical drop size. We will see later that the interface affects the turbulent fluctuations considerably at scales $r \lesssim \ell$.

The change of fluctuations in time is derived using the conditions $\delta_r u \sim U(t)$ and $\delta_r \theta \sim 1$ at the scales $r$ comparable to the size of the mixing layer $L(t)$. This yields~\cite{chertkov2003phenomenology}
	\begin{equation} 
	\label{eq5_1}
	\begin{array}l
	\displaystyle
	\delta_ru \sim U(t) \left( \dfrac{r}{L(t)}\right)^{3/5}  \sim 
	(\mathcal{A}g)^{2/5}\,\frac{r^{3/5}}{t^{1/5}},
	\\[15pt]
	\displaystyle
	\delta_r \theta \sim \left( \dfrac{r}{L(t)}\right)^{1/5}  \sim 
	(\mathcal{A}g)^{-1/5}\,\frac{r^{1/5}}{t^{2/5}}, 
	\end{array}
	\end{equation} 
where we used relations (\ref{eq4_1}). Note that these scaling laws are only approximate due to the expected intermittency~\cite{boffetta2017incompressible}. The scale $r \sim \eta(t)$ at which viscous and nonlinear terms become comparable is found as $\nu (\delta_r u)/r^2 \sim (\delta_r u)^2/r$. With the use of (\ref{eq5_1}), this yields~\cite{chertkov2003phenomenology}
	\begin{equation} 
	\label{eq5_2}
	\eta(t) \sim \frac{\nu^{5/8}}{(\mathcal{A}g)^{1/4}}\,t^{1/8}.
	\end{equation} 
In our simulation, the viscous scale computed by expression (\ref{eq5_2}) stays close to the value $\eta \approx 4$ (four lattice distances) at all times corresponding to turbulent mixing. 

\begin{figure}
	\centering
	\begin{minipage}[t]{.47\linewidth}
		\includegraphics[width=0.69\textwidth]{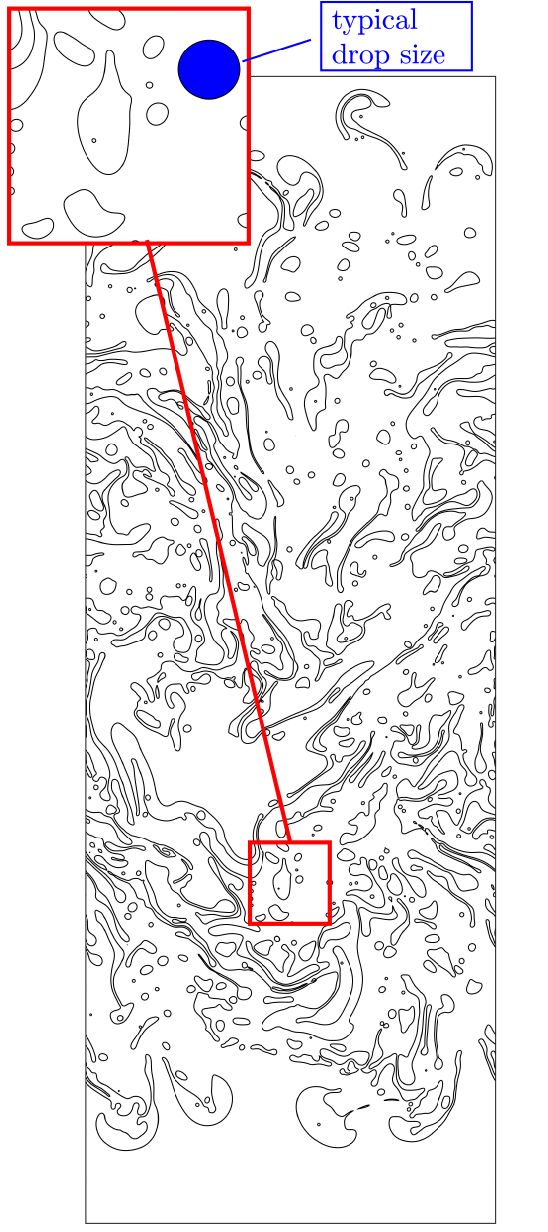}%
		 \put(-70,350){(a)}
	\end{minipage}%
	 \put(110,343){(b)}
	 \put(110,176){(c)}
	\begin{minipage}[b]{.45\linewidth}
		\includegraphics[width=0.89\textwidth]{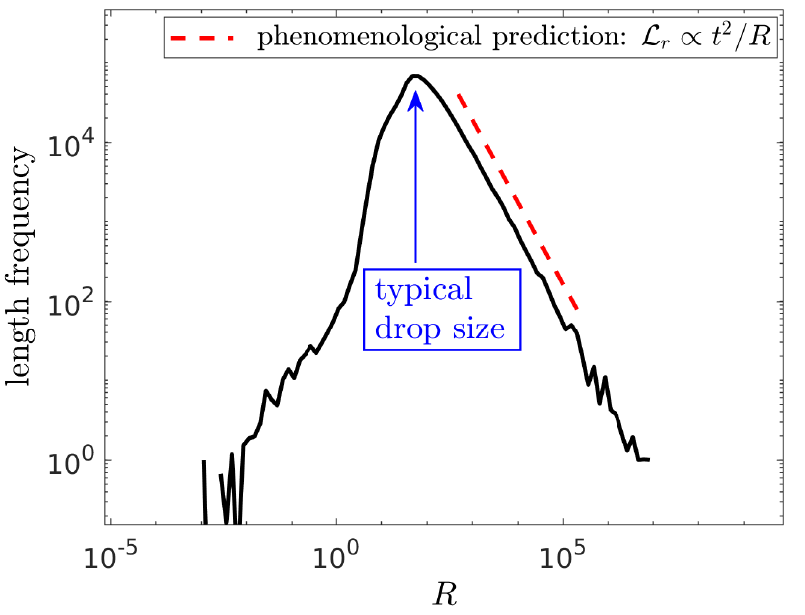}\\
	    \vspace{0.8cm}
		\includegraphics[width=0.87\textwidth]{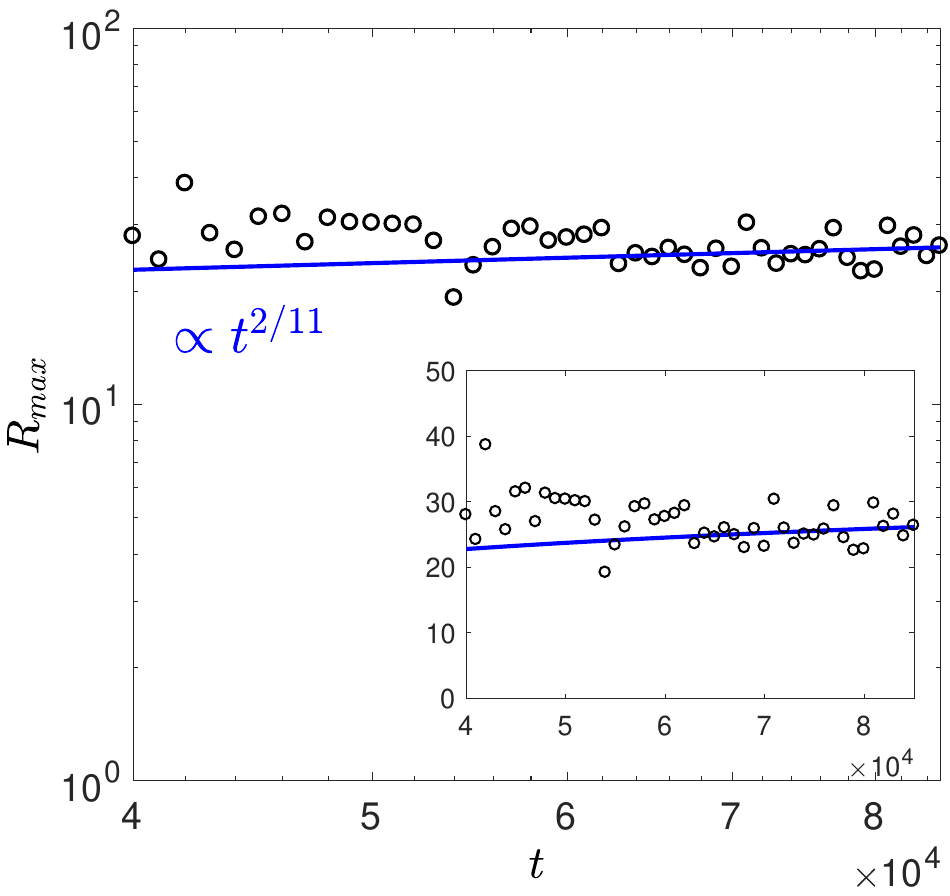}%
		\label{Loglog total length}
	\end{minipage}%
	\caption
	{%
		(a) Interface between two phases defined as the line of equal component densities, $\rho_A = \rho_B$, for a typical simulation of immiscible RT turbulence. The inset compares typical drops and their statistical size estimate (blue circle). (b) Length frequencies for different values of curvature radius $R$ along the whole interface at a fixed time. We use the logarithmic binning, which corresponds to constructing the PDF for $\log R$. The PDF maximum determines a typical drop size as $\ell = 2 R_{\max}$. The dashed red line corresponds to the theoretical prediction (\ref{eq5_5}) for the dependence of interface structures on scale, i.e., $\mathcal{L}_r \propto 1/R$.  
(c) Temporal dependence of the typical curvature radius for times corresponding to turbulent mixing, shown in logarithmic scales; the inset shows the same graph in linear scales. The blue line corresponds to the theoretical prediction $\ell = 2R_{\max} \propto t^{2/11}$,  which is expected to approximate the data for times bigger than $t \simeq 50000$, corresponding to the turbulent regime.}
\label{fig7}
\end{figure}

Let us denote by $\ell$ the size of a typical drop (or the typical size of small interface structures) in the emulsion-like state; see Fig.~\ref{fig7}(a). It can be estimated as the scale where kinetic and surface energy densities are of the same order, $\rho_0(\delta_\ell u)^2\sim \sigma/\ell$~\cite{chertkov2003phenomenology,perlekar2012droplet}. Using (\ref{eq5_1}), we find
	\begin{equation}
	\label{eq5_3}
	\ell(t) \sim \frac{\sigma^{5/11}}{\rho_0^{5/11}(\mathcal{A}g)^{4/11}}\,t^{2/11}.
	\end{equation}
This formula is derived under the assumption that the typical drop size $\ell(t)$ exceeds the viscous scale $\eta(t)$ given by expression (\ref{eq5_2}). As we  show later in Fig.~\ref{fig7}(c), a typical drop size in our simulations is about $\ell \sim 50$, which is an order of magnitude larger than the viscous scale. Therefore, $\ell$ belongs to the inertial interval at times corresponding to turbulent mixing.

If typical-sized drops are dense (distances among drops are comparable to their sizes) in the mixing layer of width $L(t)$ and horizontal length $L_x$, the total number of drops is estimated as $\mathcal{N}_\ell(t) \sim L_xL(t)/\ell^2(t)$. This yields an estimate for the maximum total length of the interface as $\mathcal{L}_{\textrm{tot}}(t) \sim \mathcal{N}_\ell(t)\ell(t) \sim L_xL(t)/\ell(t)$. Using relations (\ref{eq4_1}) and (\ref{eq5_3}), we obtain 
	\begin{equation}
	\label{eq5_4}
	\frac{\mathcal{L}_{\textrm{tot}}(t)}{L_x} \sim 
	\frac{\rho_0^{5/11}(\mathcal{A}g)^{15/11}}{\sigma^{5/11}}\,t^{20/11}.
	\end{equation}
This expression provides, up to a dimensionless coefficient, a phenomenological estimate for the growing length of the interface.

At smaller scales, the mean kinetic energy is insufficient for forming a drop. Therefore, drops of sizes $r \ll \ell$ are very rare, being induced by extreme velocity fluctuations. On the contrary, drops can form freely at larger scales $r \gg \ell$. Let us denote by $\mathcal{N}_r$ the total number of drops having size of order $r$. It is estimated similarly to typical-sized drops as $\mathcal{N}_r(t) \sim L_xL(t)/r^2$. The total interface of such drops, $\mathcal{L}_r(t) \sim \mathcal{N}_r(t)r$, is expressed using relations (\ref{eq4_1}) as
	\begin{equation}
	\label{eq5_5}
	\frac{\mathcal{L}_r(t)}{L_x} \sim \mathcal{A}g\,
	\frac{t^2}{r}.
	\end{equation}
Naturally, this length decreases for larger $r$, and, therefore, the total length of the interface is dominated by drops of typical size $r \sim \ell$.

  In the numerical simulations, the points of the moving interface $\Gamma(t)$ for an immiscible binary mixture are commonly given by the equation $\phi(\*x,t)=0$; see Fig.~\ref{fig7}(a). This definition assumes  a diffuse interface~\cite{kruger2017lattice,anderson1998diffuse} and approximates the actual interface in the sharp interface formulation given by (\ref{eq2_1}) and (\ref{eq2_2}). Then, the typical drop size can be accessed through the measurements of the interface curvature radius $R = 1/\kappa$, the inverse of the curvature $\kappa$. Therefore, we can define the typical drop size as two times the most frequent curvature radius. This concept was implemented numerically: we computed the curvature radius for each adjacent pair of small interface segments at a given time $t$, and also associated weight using the lengths of the corresponding interface segments. Then, these data are represented in the form of a histogram with logarithmic binning for the curvature radius $R$; see Fig.~\ref{fig7}(b). This histogram approximates the (not normalized) probability density function (PDF) for the values of $\ln{R}$ within the interface. The histogram in Fig.~\ref{fig7}(b) has the well-defined maximum at $R = R_{\max}(t)$, and we define the typical drop size as $\ell(t) = 2R_{\max}(t)$. The measured value is demonstrated in the inset of Fig.~\ref{fig7}(a) by a blue circle of diameter $\ell$, providing a visual validation of our numerical approach.
Figure~\ref{fig7}(c) presents the measurements of typical drop sizes at different times shown in logarithmic scale, with the straight line corresponding to the phenomenological prediction (\ref{eq5_3}). In addition to having a good agreement between theory and numerical simulations, we are able to estimate the dimensionless pre-factor in the expression (\ref{eq5_3}) as $6.7 \pm 0.7$.
Notice also that the slope of the histogram in Fig.~\ref{fig7}(b) to the right of the maximum value (dashed red line) confirms our prediction (\ref{eq5_5}) for the distribution of drops with respect to their size. This slope extends to the integral-scale structures with $R \sim L(t) \sim 10^4$. At larger values of $R\gtrsim 10^5$, Fig.~\ref{fig7}(b) measures the increased probability of  almost flat interfaces segments; such segments can be recognized both in Figs.~\ref{fig1} and~\ref{fig7}(a).

\begin{figure}
\includegraphics[width=0.4\textwidth]{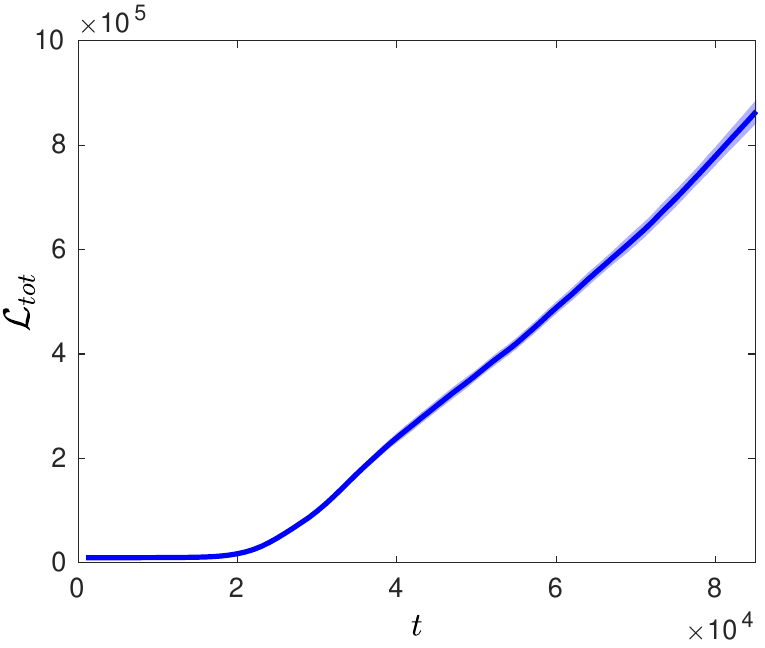}%
\put(-90,157){(a)}
\put(125,157){(b)}
\hspace{0.9cm}
\includegraphics[width=0.4\textwidth]{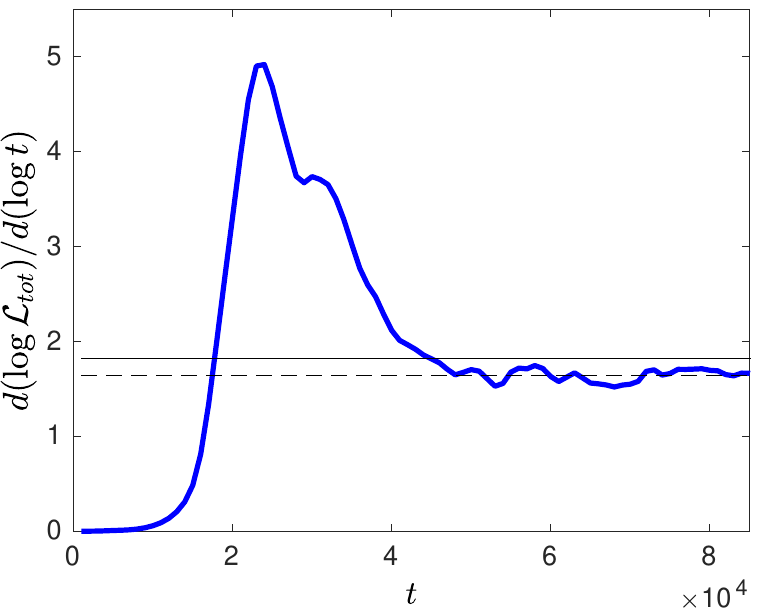}\label{Logderivative total length}
\caption{(a) Time dependence for the total interface length $\mathcal{L}_{\textrm{tot}}$ averaged over an ensemble of 10 immiscible RT simulations; the shaded region shows standard deviations. (b) Logarithmic derivative of the previous graph, $d(\log \mathcal{L}_{\textrm{tot}})/d(\log t)$, indicating the power-law dependence in the turbulent regime ($t \gtrsim 4.5\times 10^4$) with the exponent $1.64\pm0.07$ shown by a dashed horizontal line. The solid horizontal line shows the phenomenological estimate (upper bound) $20/11$ for the same exponent.}
\label{fig8}
\end{figure}

Figure~\ref{fig8}(a) presents the temporal dependence of the total interface length in our simulations, which is computed using the Cauchy--Crofton formula~\cite{do2016differential,legland2007computation}. Its logarithmic derivative (with logarithms to the base 10) is shown in Fig.~\ref{fig8}(b), demonstrating a well-established power law in the regime of turbulent mixing. The measured exponent of this power law is equal to $1.64\pm0.07$ (dashed horizontal line), which is rather close to and slightly below its theoretical estimate of $20/11$ (solid horizontal line) from Eq.~(\ref{eq5_4}). The difference between these exponents may be attributed to our theoretical assumption that typical-sized drops are dense in the mixing layer. The lower numerical value of the exponent implies that typical-sized drops get more sparse at larger times.

In the final part of this section, we study the influence of the interface on the properties of the flow. Namely, we will show that the immiscible RT turbulence generates a considerably larger enstrophy compared to the miscible flow, and that the source of this extra enstrophy is confined within a small neighborhood of the interface.

\begin{figure}
\hspace{-0.3cm}
\includegraphics[width=0.4\textwidth]{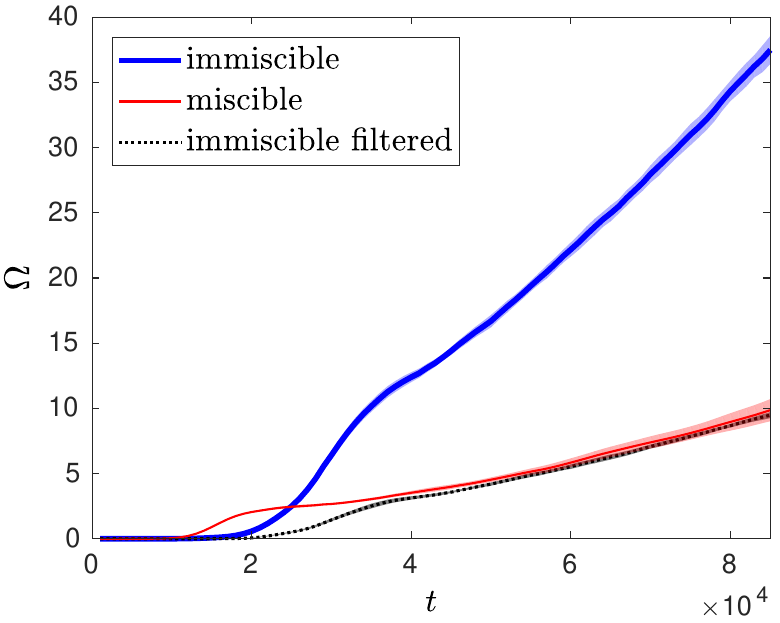}
\put(-90,160){(a)}
\put(130,160){(b)}
\hspace{1.4cm}
\includegraphics[width=0.4\textwidth]{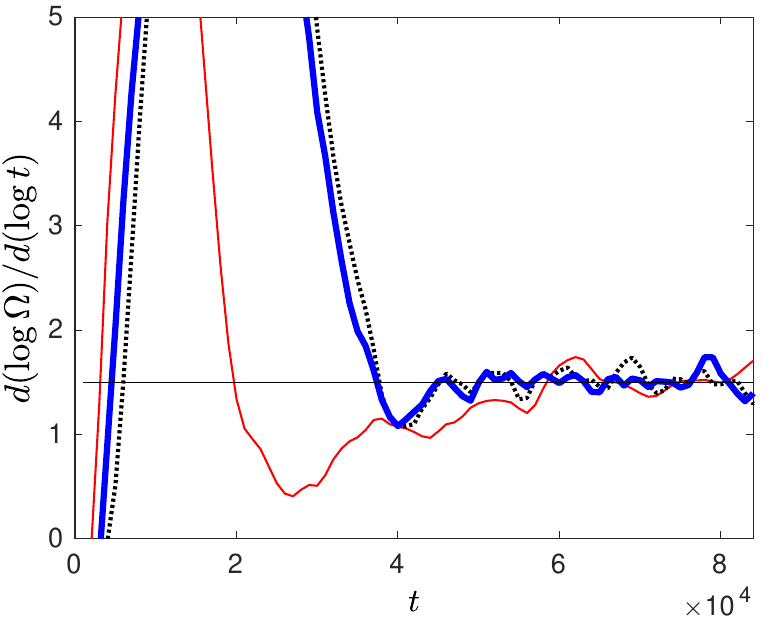} \\
\vspace{0.3cm}
\hspace{0.5cm}
\includegraphics[width=0.45\textwidth]{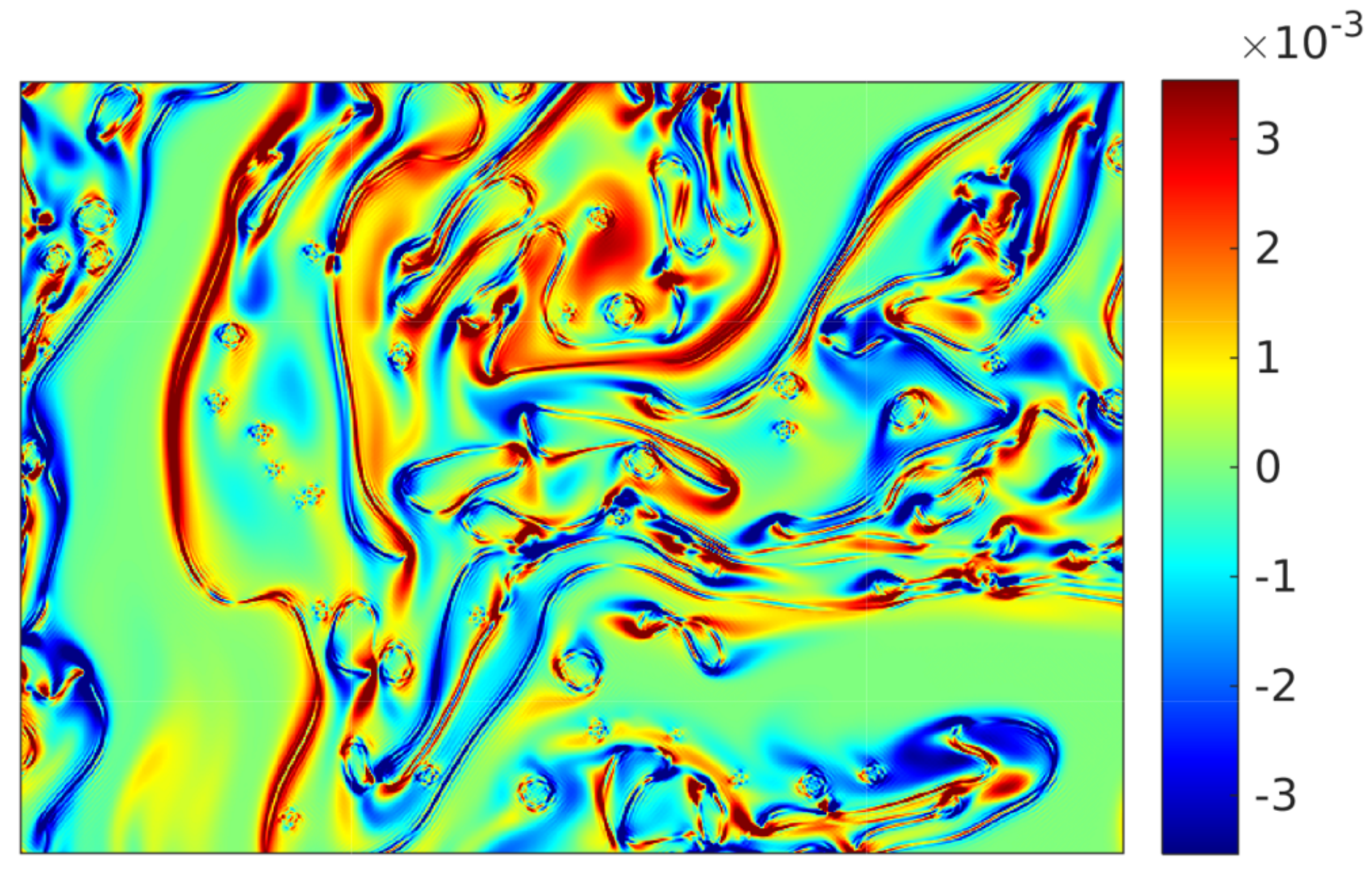}
\put(-120,140){(c)}
\put(100,140){(d)}
\hspace{0.1cm}
\includegraphics[width=0.45\textwidth]{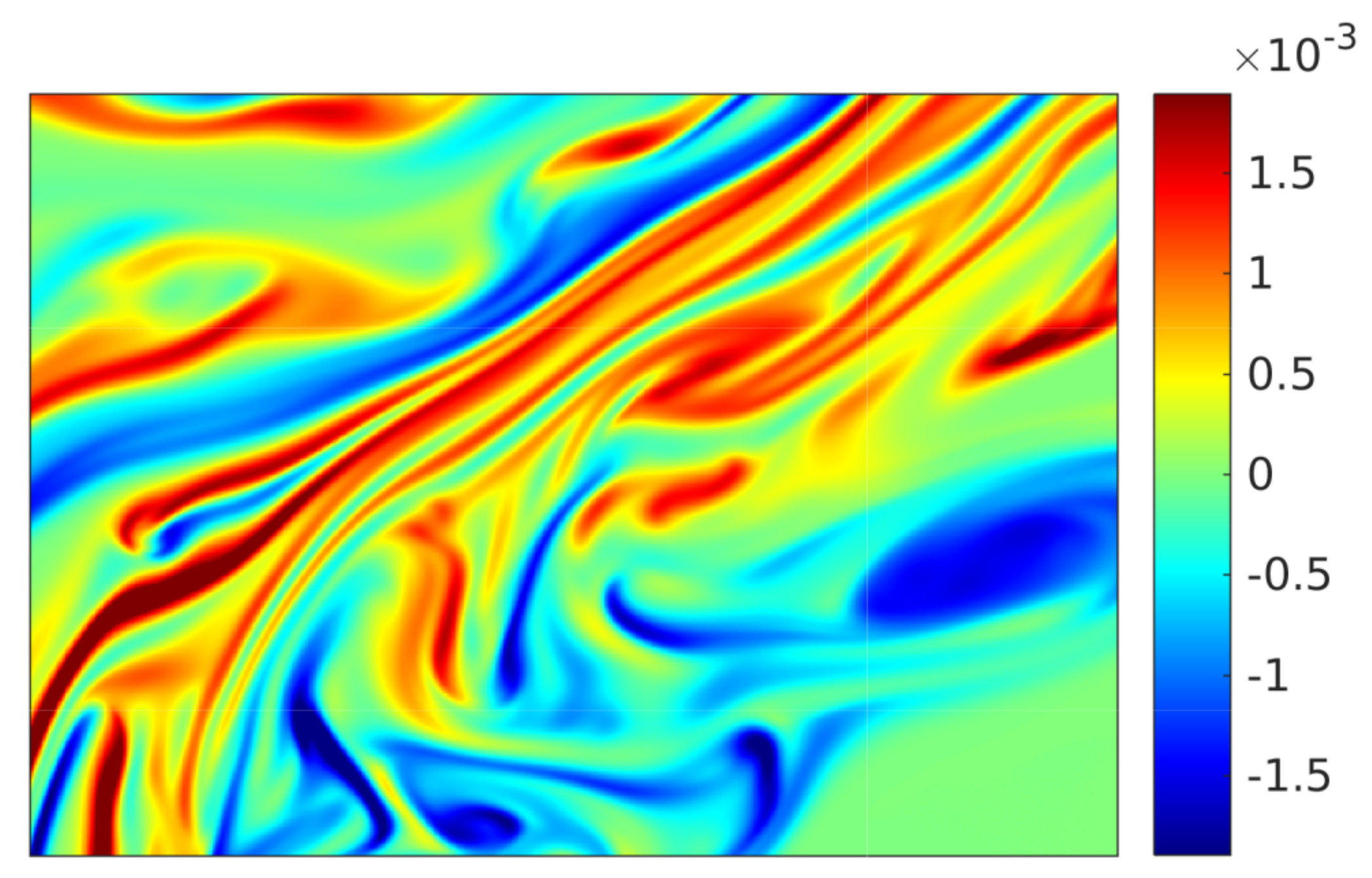}
\\
\vspace{0.5cm}
\hspace{-0.5cm}
\includegraphics[width=0.42\textwidth]{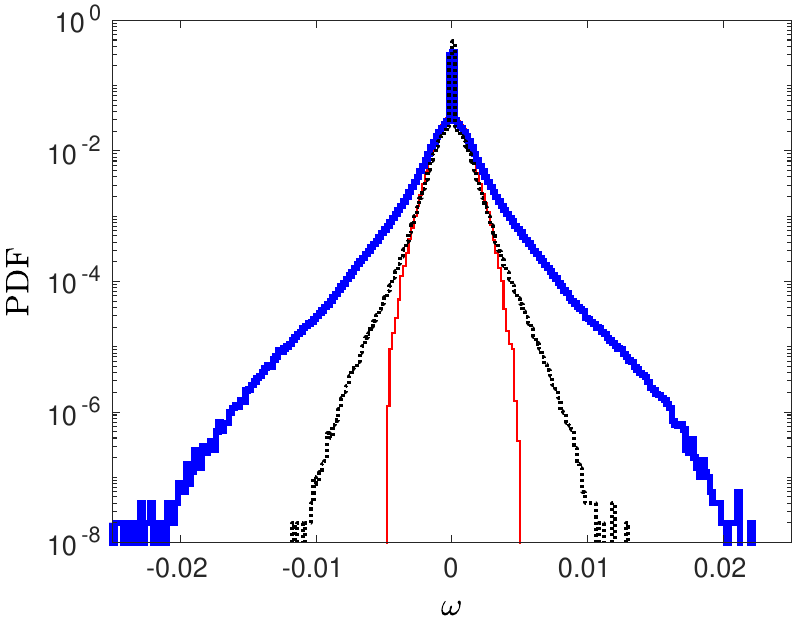}
\put(-90,165){(e)}
\put(130,165){(f)}
\hspace{1.4cm}
\includegraphics[width=0.42\textwidth]{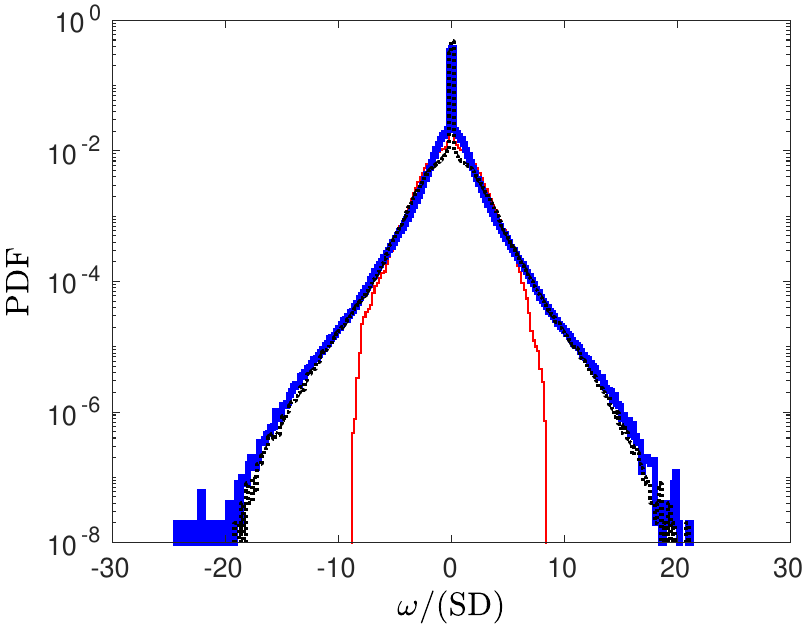}
\caption{(a) Evolution of total enstrophy averaged over $10$ realizations for the  immiscible (bold blue) and miscible (thin red) simulations; shaded regions indicate standard deviations. The dashed black line corresponds to the filtered enstrophy of the immiscible flow, by excluding small neighborhoods of the interface. (b) Logarithmic derivatives, $d(\log\Omega)/d(\log t)$, of the same graphs compared with the theoretical power law exponent (horizontal line). (c) Example of vorticity field for immiscible and (d) miscible flow. 
(e) PDFs of the vorticity fields. (f) PDFs of the vorticity fields normalized by the respective standard deviations (SD).}
\label{fig9}
\end{figure}

The phenomenological estimate for fluctuations of vorticity $\omega = \nabla \times \mathbf{u}$ in the inertial range is obtained using expression (\ref{eq5_1}) as
	\begin{equation} 
	\label{eq5_6}
	\delta_r \omega \sim \frac{\delta_r u}{r} \sim
	\frac{(\mathcal{A}g)^{2/5}}{r^{2/5}t^{1/5}}.
	\end{equation} 
Vorticity fluctuations increase at smaller scales and attain the maximum at the viscous scale $r \sim \eta(t)$. Thus, the total enstrophy of the flow $\Omega(t)$ can be estimated as a product of $(\delta_\eta \omega)^2$ and the size of the mixing layer $L(t)L_x$. Using expression (\ref{eq4_1}) for $L(t)$ and (\ref{eq5_2}) for $\eta(t)$, we derive the power law for the enstrophy $\Omega$ in the form
	\begin{equation}
	\label{eq5_7}
	\frac{\Omega}{L_x} \sim (\delta_\eta \omega)^2L(t) \sim \dfrac{(Ag)^2}{\nu^{1/2}} \,t^{3/2}.
	\end{equation}
Numerical verification of this relation is presented in Figs.~\ref{fig9}(a,b). In the first figure, we plot the total enstrophy as a function of time for the immiscible (bold blue) and miscible (thin red) flows, and the second figure shows their logarithmic derivatives demonstrating a good agreement with the phenomenological exponent $3/2$ (a horizontal line). Note that $\nu \approx 0.01$ and $D \sim 0.002$ in our miscible simulations, which implies that the particle diffusion does not affect the inertial range.

It is apparent from Fig.~\ref{fig9}(a) that, despite the power-law exponents being the same in both immiscible and miscible cases, the dimensionless pre-factor is considerably larger for the immiscible flow. We now argue that this difference can be attributed to the flow in a small neighborhood of the interface. Figure~\ref{fig9}(c) shows the vorticity field for the immiscible flow; it corresponds to a small area of $667 \times 467$ lattice points marked by the rectangle in the center of Fig.~\ref{fig1} and amplified in its right small panel. Visually, it is clear that a considerable part of the high vorticity is concentrated near the interface. For comparison, we present the vorticity field for the miscible case in Fig.~\ref{fig9}(d), which corresponds to a small area from Fig.
~\ref{fig2}. In the miscible case, the vorticity is more dispersed and its amplitude is roughly twice as small (notice the difference in the color scales). 

According to \cite{brons2014vorticity}, the interface can be considered a source of vorticity depending on the velocity jump across the interface, variations of the curvature, and other details of the flow. Also, a part of the enstrophy may have a numerical origin coming from spurious currents of the lattice Boltzmann method (see Sec.~\ref{sec3}); however, our estimates suggest that this numerical contribution is not very large~\cite{tavares2021}.
For quantification of the interface contribution, we separate the bulk enstrophy in the immiscible case by excluding small areas around the interface. This is done numerically by removing all nodes within squares of size $8 \times 8$ at each point of the interface. This size is much smaller than the typical drop ($\ell \sim 50$) and roughly twice as larger as the viscous scale ($\eta \sim 4$) and the numerical interface width ($l_{\textrm{int}} \sim 3$). The filtered enstrophy is plotted in Fig.~\ref{fig9}(a) by a dotted black curve, which agrees very well with the miscible data for the times corresponding to turbulent mixing. Though such a fine agreement may partially be attributed to the chosen filter, removing larger areas around the interface yields only a moderate effect. This observation suggests that the immiscible flow in the regions away from the interface features turbulent statistics similar to the miscible flow. This conclusion is further justified in Fig.~\ref{fig9}(e), where we plot PDFs of vorticity: one can see that the PDFs for the miscible (red) and filtered immiscible (dotted black) flows are very close, while the PDF for the full immiscible flow favors much larger values of vorticity characteristic of thin boundary layers. Still, normalized PDFs of vorticity shown in Fig.~\ref{fig9}(f) reveal a distinctive shape of the tails for large $\omega$ (rare events), which is the same for the original and filtered fields in the immiscible flow.

It is remarkable that the filtered part of the enstrophy, which is concentrated in a thin neighborhood of the interface, follows the same power law as its bulk value, Fig.~\ref{fig9}(b). We conjecture, however, that this similarity is coincidental, because the vorticity generation by the interface is not described by the Bolgiano--Obukhov scenario. The enstrophy corresponding to the interface can be estimated as a product of the total interface lengths and the linear enstrophy density. The former grows as a power law with the measured exponent $1.64\pm0.07$; see Fig.~\ref{fig8}(b). The latter may depend on the drop size and velocity fluctuations, both of which change very slowly in time;  see Eqs.~(\ref{eq5_1}) and (\ref{eq5_3}). These estimates suggest that a power law for the enstrophy growth generated by the interface may have an exponent close to $3/2$, i.e., very similar to the prediction (\ref{eq5_7}) following from the Bolgiano--Obukhov theory. Since the scaling range accessed by our simulations is not too wide, one cannot exclude other behaviors, e.g., the possibility of  anomalous scaling.

\section{Conclusion}
We have presented the a high resolution study of immiscible RT turbulence in 2D using the Shan-Chen multicomponent method. The large-scale statistics for the mixing layer, typical velocity, and average density profile have been compared with the  miscible case and found to have very similar power-law behaviors with close overall prefactors but different transient behavior. In the immiscible case, the presence of the interface affects the small-scale statistics, leading to a significant difference, with respect to the miscible RT, in the evolution of the enstrophy. The Bolgiano--Obukhov assumption generates a valid prediction for the power law behavior of the temporal evolution of total enstrophy also for the immiscible case [see Eq.~\ref{eq5_7}], but does not account for the big change in the prefactor, which could be affected by  extra vorticity induced by the interface.  
The evolution of the typical drop size and the total length of the interface in the emulsion-like state of developed RT turbulence are measured and shown to be compatible with our phenomenological predictions. 

A natural question that can be addressed in the future is about the statistics of the structures with a typical size smaller then the typical drop size. In this range of scales, the presence of capillary waves propagating along the interfaces of the drops is expected \cite{chertkov2005effects}. The developed numerical scheme can also be applied to the problem of  fragmentation and whitecapping at the surface of breaking waves, which involves a complex process with the formation of drops and bubbles; see, e.g., \cite{dyachenko2016whitecapping,mailybaev2017explosive}. 
It is also important to note that most of the numerical procedures presented in this article is naturally extendable for the three-dimensional immiscible  Rayleigh-Taylor turbulence, which is a more suitable configuration for experimental procedures, although such an extension of the present GPU code, with appropriate optimizations to obtain affordable statistics, can be a non-trivial task. Some laboratory experiments for the two-dimensional case  may be conducted in  thin liquid films~\cite{Zhou20171,carles2006rayleigh} using, for example, aqueous gelatin solutions with very high concentration~\cite{meshkov2019group}. The corresponding extension of the lattice Boltzmann method to these cases seems feasible, but it requires further study.

We thank Francesca Pelusi, from the University of Rome Tor Vergata, for the useful discussions on the technical details of the implementation of the Shan-Chen multicomponent method. We also thank Sergio Pilotto and Daniel Lins de Albuquerque for their help in some aspects of the parallel implementation of the lattice Boltzmann method on GPUs. H.T. and A.A.M. acknowledge the support from the ERC-ADG NewTURB project during their visits to the University of Rome -- Tor Vergata. AAM is supported by CNPq Grants No. 303047/2018-6 and No. 406431/2018-3. This project has received funding from the European Research Council (ERC) under the European Union’s Horizon 2020 research and innovation programme (grant agreement No 882340).

\bibliographystyle{apsrev4-1}
\bibliography{Bibliografia}

\end{document}